\documentclass[aps,prd, twocolumn, amsmath,amssymb,showpacs,eqsecnum,nofootinbib]{revtex4}

\usepackage{graphicx}
\usepackage{dcolumn}
\usepackage{bm}

\begin{document}

\title{Extended Green-Liouville asymptotics\\ and vacuum polarization for lukewarm black holes}
\author{Cormac Breen}
\email{cormac.breen@ucd.ie}
\author{Adrian C. Ottewill}
\email{adrian.ottewill@ucd.ie}
\affiliation{School of Mathematical Sciences and Complex \& Adaptive Systems Laboratory, University College Dublin, Belfield, Dublin 4, Dublin, Ireland}


\begin{abstract}
We consider a quantum field on a lukewarm black hole spacetime. We introduce a new uniform approximation to the radial equation, constructed using an extension of Green-Liouville asymptotics. We then use this new approximation to construct the renormalized vacuum polarization in the Hartle-Hawking vacuum. Previous calculations of the vacuum polarization rely on the WKB approximation to the solutions of the radial equation, however the nonuniformity of the WKB approximations obscures the results of these calculations near both horizons. The use of our new approximation eliminates these obscurities, enabling us to obtain explicitly finite and easily calculable values of the vacuum polarization on the two horizons.

\end{abstract}


\maketitle

\section{Introduction}

The properties of the quantum state in the exterior region of a nonextremal static, asymptotically flat, spherically symmetric  black hole
in thermal equilibrium at its Hawking temperature is by now well understood \cite{Candelas:1984pg,Anderson:1990jh, Anderson:1994hg, JensenOttewill:89,JensenOttewill:95,Howard:1984qp}.
However, fundamental new issues arise in the investigation of space-times with multiple horizons 
having equal or unequal surface gravity/Hawking temperatures or indeed multiple coincident horizons with zero Hawking temperature.
 Multiple horizons arise, for example, as cosmological horizons in spacetimes with a nonzero cosmological constant or as inner horizons in 
space-times with charge or spin. 
In this paper we consider a scalar quantum field on the exterior region of a Reissner-N\"{o}rdstrom-de Sitter 
black hole spacetime, where the event and cosmological horizon have the same temperature, a so-called lukewarm black hole~\cite{Romans}.

One of the simplest measures of the quantum state of a scalar field  is the expectation value 
of the square of the field, $\langle \hat{\varphi}^{2}\rangle$, the \textit{vacuum
polarization}, which is of direct physical significance to the theory of spontaneous
symmetry breaking ( see e.g., \cite{FrolovNovikov:98}).
A central difficulty in analyzing $\langle \hat{\varphi}^{2}\rangle$
arises from ultraviolet divergences which need to be renormalized to
obtain a physically meaningful quantity. In this paper we present a new approach to the calculation of  $\langle
\hat{\varphi}^{2}\rangle_{ren}$ for a quantum field on a  lukewarm black-hole spacetime  in its Hartle-Hawking state representing 
the field in thermal equilibrium at its Hawking temperature. 
(The calculation of renormalized expectation value of the stress-energy tensor, $\langle \hat{T}_{\mu\nu}\rangle_{ren}$, will be presented elsewhere \cite{BOWY}.) 
We note that once the renormalized expectation values in a particular state have been calculated,  the corresponding results in another state are easily computed,
as the difference between expectation values in two states does not require renormalization ,as the ultraviolet divergences are geometrical and state-independent.

Previous calculations of  $\langle\hat{\varphi}^{2}\rangle_{ren}$ for lukewarm black holes~\cite{Winstanley:2007} make use of the WKB approximation. This method encounters significant difficulties near the horizons of the black hole, as will be outlined in Sec.~\ref{sec:PreviousResults}. We introduce a new uniform approximation, constructed using an extension of Green-Liouville asymptotics due to Olver~\cite{Olver}, which over comes these difficulties and enables us to construct  $\langle
\hat{\varphi}^{2}\rangle_{ren}$  for lukewarm black holes in a more satisfactory manner.

Other approximations have been developed which are uniform in $l$ near the horizons~\cite{Tom}.  While these approximations can be used to prove the regularity of $\langle \hat{\varphi}^2 \rangle_{ren}$ on the horizons of a lukewarm black hole~\cite{Winstanley:2007}, they suffer from two major drawbacks. First, they fail to give exact horizon values for $\langle \hat{\varphi}^2 \rangle_{ren}$ for all parameter sets. Second, and perhaps most significantly, they cannot be used to investigate the regularity of $\langle T_{\mu \nu} \rangle_{ren}$  on the black-hole horizons.
Our new approximation overcomes these shortcomings which, in this paper, will allow us to obtain explicitly finite and easily calculable values of $\langle \hat{\varphi}^2 \rangle_{ren}$  on the two horizons and  in a future paper~\cite{BOWY}  will provide a route for us to extend this analysis to $\langle T_{\mu \nu} \rangle_{ren}$.

Frolov et al.~\cite{Frolov} have developed an approximation for 2-D static black-hole spacetimes which is valid  in the entire exterior region. However, our approximation is constructed in a systematic manner which affords us greater control over the error of the approximation, a property which is key for the extension of our analysis to $\langle T_{\mu \nu} \rangle_{ren}$.

This paper is organized as follows. In Sec.~\ref{sec:PreviousResults} we will outline the existing results and the issues they encounter. In Sec.~\ref{sec:GreenLiouville} we introduce Olver's extended Green-Liouville (EGL) asymptotics and demonstrate its application to this particular calculation. In Secs.~\ref{sec:FormalCalculation}, \ref{sec:NumericalCalculations}, and \ref{sec:HorizonValue} we describe  the calculations needed for $\langle
\hat{\varphi}^{2}\rangle_{ren}$, including a new method for obtaining explicitly finite horizon values. Sec.~\ref{sec:Plots} contains plots of our results. Finally our conclusions are presented in Sec.~\ref{sec:Conclusions}. Throughout this paper we use units in which $8\pi G=\hbar =c =k_B =1$.

\section{Previous Results}
\label{sec:PreviousResults}

\subsection{ Construction of $\langle
\hat{\varphi}^{2}\rangle_{ren}$}

We will briefly outline the standard method used for the calculation of $\langle
\hat{\varphi}^{2}\rangle_{ren}$; a more detailed description can be found in \cite{Anderson:1994hg}. We begin by considering a scalar field 
$\hat{\varphi}$ with mass $m$ and coupling  $\xi$ to the Ricci scalar $R$. The field satisfies
\begin{equation}
(\Box-m^2-\xi R)\hat{\varphi} =0.
\end{equation}

We will perform our calculations on a static, spherically symmetric background spacetime with line element:
\begin{equation}
\label{le}
\mathrm{d}s^2=-f(r) \mathrm{d} t^2 +\frac{1}{f(r)} \mathrm{d} r^2 +r^2 \mathrm{d}\theta^2 + r^2 \sin^2 \theta \mathrm{d}\phi^2 .
\end{equation}
The spacetime will have a horizon at $r=r_0$ whenever $f(r_0)=0$ and in this case the surface gravity of that horizon is given by $\kappa_0=|f'(r_0)|/2$ .

Following the standard procedure we Euclideanize our spacetime, that is we perform a Wick rotation $\tau \to i t$, Eq.~(\ref{le}) then becomes
\begin{equation}
\label{lee}
\mathrm{d}s^2=f(r) \mathrm{d} \tau^2 +\frac{1}{f(r)} \mathrm{d} r^2 +r^2 \mathrm{d}\theta^2 + r^2 \sin^2 \theta \mathrm{d}\phi^2.
\end{equation}
Assuming $\kappa_0 \neq 0$, this space will have a conical singularity whenever $f(r_0)=0$ which may be removed by making $\tau$ periodic with period $2\pi/\kappa_0$.
This periodicity in the Euclidean section corresponds in quantum field theory on the Lorentzian section to a thermal state with temperature
$T=\kappa_0/2\pi$.

The unrenormalized vacuum polarization can now be written as the coincidence limit of the two point Euclidean Green's function, which satisfies
\begin{equation}
(\Box_x-m^2-\xi R)G_E(x,x') =-\frac{\delta^4(x,x')}{\sqrt{g(x)}},
\end{equation}
so we have
\begin{equation}
\langle\hat{\varphi}^2\rangle_{unren}=\lim_{x\to x'}G_E(x,x').
\end{equation}
We are taking our field to be in a Hartle-Hawking state, corresponding to the Eucidean solution with the conical singularity at $r_0$ removed, as such $G_E(x,x')$  takes the form~\cite{Anderson:1994hg}
\begin{align}
\label{genps}
&G_E(x,x') =\frac{T}{4 \pi}\sum_{n=-\infty}^{\infty} \mathrm{e}^{ i n \kappa_0 (\tau-\tau')}\nonumber\\
&\quad\times\sum^{\infty}_{l=0}(2l+1) P_l(\cos\gamma)C_{n l}p_{n l}(r_<)q_{n l}(r_>) ,
\end{align}
where $P_l$ is the Legendre polynomial and $\cos\gamma= \cos\theta\cos\theta' +\sin\theta \sin\theta'\cos(\varphi-\varphi')$. The functions $p_{n l}$ and $q_{n l}$ satisfy the differential equation
\begin{align}
\label{mode}
 &\frac{d^2 S}{dr^2} +\bigg(\frac{2 f}{r}+\frac{df}{dr}\bigg) \frac{d S}{dr}\nonumber\\
 &\quad-\bigg(\frac{n^2 \kappa_0^2}{f} +\frac{l(l+1)}{r^2} +m^2 +\xi R\bigg) S=0,
\end{align}
with appropriate boundary conditions, which will be discussed in Sec.~\ref{sec:NumericalCalculations}. This equation cannot, in general, be solved analytically and so $p_{n l}$ and $q_{\omega l}$ have to be obtained numerically. The normalization constant $C_{n l}$ is fixed by the Wronskian condition
\begin{equation}
\label{Wron}
C_{n l}\bigg[p_{n l}\frac{d q_{n l}}{dr} -q_{n l}\frac{d p_{n l}}{dr}\bigg] =-\frac{1}{r^2 f}.
\end{equation}
We could, of course, absorb this into the definition of the functions but it is frequently convenient to normalise  $p_{n l}$ and $q_{n l}$
by their behaviour near different singular points of Eq.~(\ref{mode}). 

The standard procedure to renormalize is to first regularise by separating $x$ and $x'$, choosing a temporal splitting as it simplifies numerical calculations. Correspondingly we set $r=r'$, $\theta=\theta'$, and $\varphi=\varphi'$ and we define $\epsilon=\tau-\tau'$. Inserting these conditions into Eq.~(\ref{genps}) we have
\begin{align}
\label{tempps}
G_E(x,x') &= \frac{T}{4 \pi}\sum_{n=-\infty}^{\infty} \mathrm{e}^{ i n\kappa_0 \epsilon} 
\sum^{\infty}_{l=0}(2l+1) C_{n l}p_{n l}(r)q_{n l}(r).
\end{align}
Due to the distributional nature of the sum, this approach carries with it a superficial divergence in the Greens's function (\ref{tempps}),
manifested through the nonconvergence of the sum over $l$ in Eq.~(\ref{tempps}). This divergence is nonphysical and occurs even when the points are separated. Fortunately, due to the high degree of symmetry here, it is possible to resolve this issue by subtracting multiples of the delta function (which vanish when the points are separated) inside the sum, which render the sum convergent. In more general space-times,
a more careful approach avoids these superficial divergences~\cite{OttewillTaylor:2010}. In the present case, the exact form of the subtraction term is well known and is given by  \cite{Anderson:1994hg} 
\begin{align}
\label{reg}
&G_E(x,x') =\nonumber\\
&\frac{\kappa_0}{4 \pi^2}\sum_{n=1 }^{\infty} \cos(n \kappa_0\epsilon)
\sum^{\infty}_{l=0}\bigg[(2l+1) C_{n l}p_{n l}(r_<)q_{\omega l}(r_>)-\frac{1}{r f^\frac{1}{2}}\bigg]\nonumber\\
&+\frac{T}{4 \pi}\sum^{\infty}_{l=0}\bigg[(2l+1) C_{0 l}p_{0 l}(r_<)q_{0 l}(r_>)-\frac{1}{r f^\frac{1}{2}}\bigg],
\end{align}
where for later convenience we have separated out the $n=0$ term.
To renormalize this expression we subtract the Christensen renormalization counterterms and then take the limit $\epsilon \to 0$. These terms are given by
\begin{align}
\label{div}
\langle\hat{\varphi}^2\rangle_{div}&= \frac{1}{8 \pi^2\sigma}+\frac{1}{16 \pi^2}\left(m^2 +\left(\xi-{\textstyle\frac{1}{6}}\right)R\right)\ln\left(\frac{\mathrm{e}^{2\gamma}\mu^2 | \sigma|}{2}\right)\nonumber\\
&\qquad-\frac{m^2}{16\pi^2} +\frac{1}{96\pi^2}R_{\alpha\beta}\frac{\sigma^{\alpha}\sigma^{\beta}}{\sigma}
\end{align}
Here $\sigma$ is equal to one half the square of the distance between the separated points along the geodesic connecting them, $\sigma^{\alpha} \equiv \sigma^{; \alpha}$, $\gamma$ is Euler's constant and $R_{\alpha\beta}$ is the Ricci tensor. For a massive scalar field the constant $\mu$ is conventionally set equal to $m$; however, for a massless scalar field the constant $\mu$ is arbitrary \cite{Anderson:1994hg}.

For temporal splitting we have
\begin{align*}
\sigma &=\frac{1}{2} f \epsilon^2 -\frac{1}{96\pi^2} f'{}^2\epsilon^4 +O(\epsilon^6),\\
\sigma^r =\frac{1}{4} ff'\epsilon^2  &+ O(\epsilon^4),\qquad
\sigma^t=-\epsilon  +\frac{1}{24\pi^2}  f'{}^2\epsilon^3 +O(\epsilon^4),
\end{align*}
while $\sigma^{\theta}=\sigma^{\phi}=0$.
$\langle\hat{\varphi}^2\rangle_{div}$ then simplifies to
\begin{align}
\label{div2}
\langle\hat{\varphi}^2\rangle_{div}&=\frac{1}{4 \pi^2 f \epsilon^2}+ \frac{1}{8\pi^2}\left(m^2 +\left(\xi-\textstyle{\frac{1}{6}}\right)R\right)\ln\left(\frac{\mathrm{e}^{2\gamma}\mu^2f\epsilon^2}{4}\right)\nonumber\\
&-\frac{m^2}{16\pi^2} +\frac{f'{}^2}{192\pi^2 f} 
  -\frac{f''}{96\pi^2 } -\frac{f'}{48\pi^2 r} . \
\end{align}
We may now express Eq.~(\ref{div2}) in terms of mode sums using the following distributional  identities, valid for small $\epsilon$ and any $\kappa_0 >0$~\cite{Howard:1985yg} ,
\begin{align*}
&\frac{1}{\epsilon^2}=\kappa_0{}^2 \sum_{n=1}^{\infty} n \cos(n \kappa_0 \epsilon) -\frac{\kappa_0{}^2}{12} + O(\epsilon^2)\\
-&\frac{1}{2}\ln(\kappa_0{}^2 \epsilon^2) =\sum_{n=1}^{\infty} \frac{\cos(n \kappa_0 \epsilon)}{n} +O(\epsilon^2).
\end{align*}
After subtracting Eq.~(\ref{div2}) from Eq.~(\ref{reg}) we may take the limit $\epsilon \to 0$ yielding the result~\cite{Anderson:1994hg} 
\begin{equation*}
\langle\hat{\varphi}^2\rangle_{ren}=\langle\hat{\varphi}^2\rangle_{numeric}+\langle\hat{\varphi}^2\rangle_{analytic}
\end{equation*}
with
\begin{align}
\label{num}
&\langle\hat{\varphi}^2\rangle_{numeric}=\frac{\kappa_0}{4 \pi^2}\sum_{n=1 }^{\infty} \bigg\{\frac{n \kappa_0}{f} +\frac{1}{2n \kappa_0}\left(m^2 +\left(\xi-\textstyle{\frac{1}{6}}\right)R\right)\nonumber\\
&\quad +\sum^{\infty}_{l=0}\left[(2l+1) C_{n l}p_{n l}(r)q_{\omega l}(r)
-\frac{1}{r f^\frac{1}{2}}\right]\bigg\}\nonumber\\
&+\frac{\kappa_0}{8 \pi^2}\sum^{\infty}_{l=0}\bigg[(2l+1) C_{0 l}p_{0 l}(r)q_{0 l}(r)-\frac{1}{r f^\frac{1}{2}}\bigg],
\end{align}
\begin{align}
\label{anal}
&\langle\hat{\varphi}^2\rangle_{analytic}= - \frac{1}{16\pi^2}\left(m^2 +\left(\xi-\textstyle{\frac{1}{6}}\right)R\right)\ln\left(\frac{\mathrm{e}^{2\gamma}\mu^2f}{ 4 \kappa_0{}^2}\right)\nonumber\\ 
&\quad +\frac{m^2}{16\pi^2} -\frac{f'{}^2}{192\pi^2 f} 
  +\frac{f''}{96\pi^2 }+\frac{f'}{48\pi^2 r} +\frac{\kappa_0{}^2}{48 \pi^2 f}. 
\end{align}

\subsection{ Previous calculations of $\langle
\hat{\varphi}^{2}\rangle_{ren}$}
\label{sec:PreviousResultsB}
The sums over $l$ and $n$ contained in Eq.~(\ref{num}) converge so slowly as to make their computation impractical. To solve this problem one makes use of an approximation to $C_{n l}p_{n l}(r)q_{\omega l}(r)$. The idea is that one subtracts this approximation inside the sum over $l$, which causes the sums over $l$ and $n$ to converge rapidly. Then one performs the sum of the approximation explicitly and adds this back onto the rapidly convergent sum. Of course, this does not affect the final answer as all one is doing is subtracting the approximation and then adding it back on again.

The standard approximation that is used is the WKB approximation and, indeed, previous calculations for the  lukewarm case adopt this approach  \cite{Winstanley:2007}. The WKB approximation, however, suffers from problems with uniformity: the closer to the horizon one gets the 
higher $l$ one requires. Thus, if one considers regions bounded away from the horizons of the space-time under consideration then the WKB approximation is useful; however, it fails to capture the correct behavior as these horizons are approached. This nonuniformity is of particular importance when one considers the case that both $\langle
\hat{\varphi}^{2}\rangle_{numeric}$ and $\langle
\hat{\varphi}^{2}\rangle_{analytic}$ diverge on the horizons. From inspection of Eqs. (\ref{num}) and (\ref{anal}) and the definition of the surface gravity 
one sees that this is the case whenever
$m^2 +(\xi-\frac{1}{6})R$
 is nonvanishing . It can be shown analytically that these divergences do in fact cancel~\cite{Winstanley:2007} however the numerical cancellation of these two divergent quantities obscures calculations  near the horizons and reduces the accuracy of the results obtained. Another consequence is that it is impossible to find explicit values for $\langle\hat{\varphi}^{2}\rangle_{ren}$ on the horizons. In addition, it is far from obvious how to extend the analytic results to derivatives of the field as required in calculations of  $\langle T_{\mu\nu}\rangle_{ren}$.

In Sec.~\ref{sec:GreenLiouville} we will show how replacing the WKB approximation by the uniform EGL approximationeliminates these obscurities enabling us to obtain explicitly finite and easily calculable values of the vacuum polarization on the two horizons.

\section{Extended Green-Liouville Asymptotics}
\label{sec:GreenLiouville}
The fundamental problem that faces us is to obtain suitable approximate solutions to 
a second order differential equation of the form
\begin{equation}
\frac{\mathrm{d}^2w}{\mathrm{d}x^2} =\left(k^2 F(x) +G(x)\right)w,
\end{equation}
 with a large parameter $k$. The form of the approximate solution of this equation depends on the zeros and singularities of $F$ and $G$
in the region under consideration. In the case where there are no zeros or singularities, Liouville~\cite{Liouville} and Green~\cite{Green} independently developed a uniform approximation to solutions of such an equation. This approximation is widely known in the physics literature as the WKB approximation in recognition of the development of the theory by Wentzel~\cite{Wentzel}, Kramers~\cite{Kramers}, and Brillouin~\cite{Brillouin}. This approximation remains valid under some loosening of these conditions on
$F$ and $G$; however, significant reanalysis is needed in the case where the region under consideration contains a point $x_0$ where $F(x)$ has a simple pole and $G(x)$ has a double pole.
This is precisely the case which arises when the radial equation~\ref{mode} is transformed into the current form.
The analysis to find uniform approximations to differential equations which possess such transition points was developed by Olver and summarised in his book~\cite{Olver}.  We will now briefly outline this analysis before demonstrating its application for the case under consideration; a more detailed description can be found in Chapter 12 of Ref~\cite{Olver}.

\subsection{Background Theory}
\label{GreenLiouvilleA}
We begin by considering solutions of equations of the form 
\begin{equation}
\label{norm}
\frac{\mathrm{d}^2w}{\mathrm{d}x^2} =\left(k^2 F(x) +G(x)\right)w,
\end{equation}
in which $k$ is a large parameter, and at $x=x_0$, say, $F(x)$ has a simple pole while $(x-x_0)^2 G(x)$ is analytic.
We  first transform our independent and dependent variables in the following manner:
\begin{align*}
\frac{1}{\xi}\left(\!\frac{\mathrm{d}\xi}{\mathrm{d}x}\!\right)^{\!\!2}=4F(x), \qquad 
w=\left(\!\frac{\mathrm{d}\xi}{\mathrm{d}x}\!\right)^{\!\!-1/2} W.
\end{align*}
We assume, without loss of generality, that $F(x)$ has the same sign as $x-x_0$. Integration then yields
\begin{align}
\label{xi}
\xi^{1/2}&=\int^{x}_{x_0} F^{1/2}(x') \mathrm{d}x' \quad (x\geq x_0)\nonumber\\
(-\xi)^{1/2}&=\int^{x}_{x_0} \left(-F(x')\right)^{1/2} \mathrm{d}x' \quad (x\leq x_0)
\end{align}
These equations determine a continuous one-to-one correspondence between the variables $x$ and $\xi$.
The transformed differential equation is given by
 \begin{equation}
 \label{GL1}
\frac{\mathrm{d}^2W}{\mathrm{d}\xi^2} =\left(\frac{k^2}{4\xi} +\hat{\psi}(x)\right)W,
\end{equation}
where
\begin{align*}
 \hat{\psi}(x)&=\frac{G(x)}{\hat{F}(x)} + \frac{1}{\hat{F}(x)^{1/4} }\frac{\mathrm{d}^2\hat{F}(x)^{1/4}}{\mathrm{d}\xi^2}\nonumber\\ 
 \hat{F}(x)&=\left(\!\frac{\mathrm{d}\xi}{\mathrm{d}x}\!\right)^{\!\!2}=4\xi F(x).
\end{align*}
If $G(x)$ has a simple or double pole at $x=x_0$, then $\hat{\psi}(x)$ has the same kind of singularity at $\xi=0$. We denote the value of $\xi^2\hat{\psi}(x)$ at $\xi=0$ by $\frac{1}{4}(n^2-1)$, and rearrange 
(\ref{GL1}) in the form
 \begin{equation}
 \label{GL2}
\frac{\mathrm{d}^2W}{\mathrm{d}\xi^2} =\left(\frac{k^2}{4\xi} +\frac{n^2-1}{4\xi^2} +\frac{\psi(x)}{\xi}\right)W,
\end{equation}
in which $\psi(x)=\xi\hat{\psi}(x)-\frac{1}{4}(n^2-1)\xi^{-1}$ and is analytic at $\xi=0$. This equation has, for each value of $k$ and each nonnegative integer $j$, solutions $W_{2j+1,1}(k,\xi)$ and $W_{2j+1,2}(k,\xi)$ which are repeatedly differentiable in the interval $(0,\beta)$, and are given by
\begin{subequations}
\begin{align}
\label{expant}
&W_{2j+1,1}(k,\xi) = \xi^{1/2}I_{n}(k\xi^{1/2})\sum^{j}_{s=0} \frac{A_s(\xi)}{k^{2s}} \nonumber\\
&\ +\frac{\xi}{k}I_{n+1}(k\xi^{1/2})\sum^{j-1}_{s=0} \frac{B_s(\xi)}{k^{2s}} +\epsilon_{2j+1,1}(k,\xi),
\end{align}
\begin{align}
\label{expantb}
&W_{2j+1,2}(k,\xi) =\xi^{1/2}K_{n}(k\xi^{1/2})\sum^{j}_{s=0} \frac{A_s(\xi)}{k^{2s}}\nonumber\\
&\  -\frac{\xi}{k}K_{n+1}(k\xi^{1/2})\sum^{j-1}_{s=0} \frac{B_s(\xi)}{k^{2s}} +\epsilon_{2j+1,2}(k,\xi),
\end{align}
\end{subequations}
where $I_n$ and $K_n$ are the modified $n$th order Bessel functions of the first
and second kind respectively. 
The coefficient $A_0(\xi)$ is a constant which, without loss of generality, we take to be 1 while the higher order coefficients $A_s(\xi)$, $B_s(\xi)$ are determined by the recursion relations:
\begin{align}
\label{recur}
&B_s(\xi)=-A_s'(\xi)  \nonumber\\
&\quad + \frac{1}{\xi^{1/2}} \int^{\xi}_{0} \{\psi(\xi')A_s(\xi')  
-\left(n+\textstyle{\frac{1}{2}}\right)A_s'(\xi')\}\frac{d\xi'}{\xi'^{1/2}},\nonumber\\
&A_{s+1}(\xi) = nB_s{\xi}-\xi B_s'(\xi) +\int \psi(\xi)B_s(\xi)d\xi.
\end{align}
The error terms  $\epsilon_{2j+1,1}(k,\xi)$ and $\epsilon_{2j,+1,2}(k,\xi)$ associated with each solution satisfy the bounds 
\begin{subequations}
\begin{align}
&|\epsilon_{2j+1,1}(k,\xi)|\leq\lambda_n\xi^{1/2}I_{n}(k\xi^{1/2})\nonumber\\
&\qquad\times\exp\left(\frac{\lambda_n}{k}\mathcal{V}_{0,\xi}(\xi^{1/2}B_0)\right)\frac{\mathcal{V}_{0,\xi}(\xi^{1/2}B_j)}{k^{2j+1}},\\
&|\epsilon_{2j+1,2}(k,\xi)|\leq\lambda_n\xi^{1/2}K_{n}(k\xi^{1/2})\nonumber\\
&\qquad\times\exp\left(\frac{\lambda_n}{k}\mathcal{V}_{\xi,\beta}(\xi^{1/2}B_0)\right)\frac{\mathcal{V}_{\xi,\xi(x_0)}(\xi^{1/2}B_j)}{k^{2j+1}},\\
&\left|\frac{\partial\epsilon_{2j+1,1}(k,\xi)}{\partial \xi}-\frac{n+1}{2\xi} \epsilon_{2j+1,1}(k,\xi)\right|\nonumber\\
&\leq\lambda_n\xi^{1/2}\frac{K_{n+1}(k\xi^{1/2})}{K_{n}(k\xi^{1/2})}I_{n}(k\xi^{1/2})\nonumber\\
&\qquad\times\exp\left(\frac{\lambda_n}{k}\mathcal{V}_{0,\xi}(\xi^{1/2}B_0)\right)\frac{\mathcal{V}_{0,\xi}(\xi^{1/2}B_j)}{k^{2j}},\\
&\left|\frac{\partial\epsilon_{2j+1,2}(k,\xi)}{\partial \xi}-\frac{n+1}{2\xi} \epsilon_{2j+1,2}(k,\xi)\right|\nonumber\\
&\leq\lambda_n\xi^{1/2}K_{n+1}(k\xi^{1/2})\exp\left(\frac{\lambda_n}{k}\mathcal{V}_{\xi,\xi(x_0)}(\xi^{1/2}B_0)\right)\nonumber\\
&\qquad\times\frac{\mathcal{V}_{\xi,\beta}(\xi^{1/2}B_j)}{k^{2j}},
\end{align}
\end{subequations}
where $\mathcal{V}_{a,b}(f)$ is the variational operator defined by
\begin{equation*}
\mathcal{V}_{a,b}(f)= \int^{b}_{a} |f'(x)|dx
\end{equation*}
and
\begin{align*}
\lambda_n=  \begin{cases} 1.07& \quad n=0\\ 1& \quad n>0 \end{cases}.
\end{align*}

\subsection{Application to black hole spacetimes}
\label{sec:GreenLiouvilleB}
It can be shown that Eq.~(\ref{mode}) can be written in the form of Eq.~(\ref{norm}). Hence we can apply the above theory to find a uniform approximation to either $p_{n l}$ or $q_{n l}$ of order $j$ to 
Eq.~(\ref{mode})  by truncating the series Eq.~(\ref{expant}) or Eq.~(\ref{expantb}) at an appropriate $j$. In the spirit of Sec.~\ref{GreenLiouvilleA}  we perform the following transformation of variables
\begin{align}
r \to \xi &=\bigg(\int_{r_0}^{r} \frac{1}{\sqrt{r'^2 f}} dr'\bigg)^2\nonumber\\
S \to W &= (\xi r^2 f)^{1/4} S,
\end{align}
where $r_0$ denotes the location of the  horizon.
Eq.~(\ref{mode}) then becomes
\begin{equation*}
\frac{d^2 W}{d \xi^2} = \bigg(\frac{k^2}{4\xi} +\frac{n^2-1}{4 \xi^2} +\frac{\psi(\xi)}{\xi}\bigg)W
\end{equation*}
with
\begin{align*}
k^2 &= l(l+1) +\left(m^2 +\left(\xi-\textstyle{\frac{1}{6}}\right)R(r_0)\right)r_0^2\nonumber\\
&\qquad +n^2\bigg(\frac{R(r_0)r_0^2}{6} +2\kappa_0 r_0\bigg) +\frac{1}{3}(1-n^2)\nonumber\\
&\equiv l(l+1) + N.
\end{align*}
and
\begin{align}
\psi(\xi)&= \frac{r^2 n^2 \kappa^2}{4f} +\frac{r^2(m^2 +\xi R(r))}{4} +\frac{f}{16} +\frac{3 r f'}{16} \nonumber\\
&+\frac{r^2 f'}{16} -\frac{r^2 f'^2}{64 f} -\frac{3}{16\xi}- \frac{n^2-1}{4\xi} -\frac{N}{4}.
\end{align}
\begin{figure}[hb]\centering
\includegraphics[width=9cm]{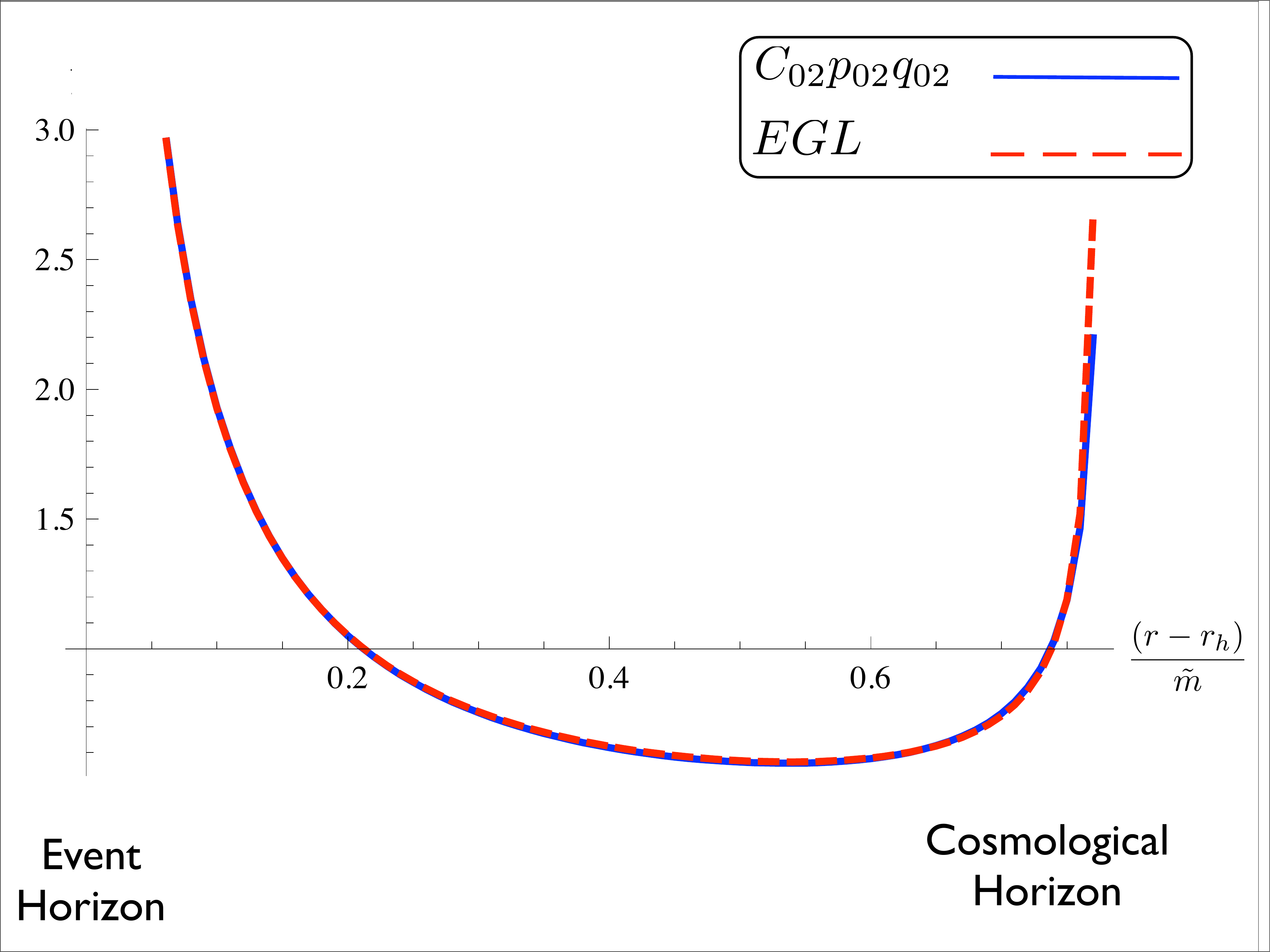}\caption{\label{fig:Compar}A comparison plot between the exact solution $C_{02}p_{02}q_{02}$ and its zeroth order EGL approximation $P_0Q_0$ for a massive conformally coupled scalar field with mass $m$ on a lukewarm blackhole with $M=Q=0.1\tilde{m}$ and $\Lambda= 3 \tilde{m}^{-2}$. The approximation can be seen to capture the  behaviour the event horizon while it the breaks down near the cosmological horizon.}
\end{figure}
It can be easily shown, using the expansions in Appendix \ref{Ap:NearHorizon}, that $\psi(\xi)$ is $O(\xi)$ as $\xi \to 0$ ($\equiv r\to r_0$).  From Sec.~\ref{GreenLiouvilleA}  we then have the following uniform $j$th order approximations for $p_{n l}$ and $q_{n l}$ respectively:
\begin{subequations}
\begin{align}
\label{expan}
P_{j}(k,r) &=\frac{\xi(r)^{1/4}}{(r^2 f)^{1/4}}I_{n}(k\xi(r)^{1/2})\sum^{j}_{s=0} \frac{A_s(\xi(r))}{k^{2s}} \nonumber\\
&+\frac{\xi(r)^{3/4}}{k(r^2 f)^{1/4}}I_{n+1}(k\xi(r)^{1/2})\sum^{j-1}_{s=0} \frac{B_s(\xi(r))}{k^{2s}}\nonumber\\
&+\tilde{\epsilon}_{2j+1,1}(k,\xi(r)),{}\\
Q_{j}(k,r) &=\frac{\xi(r)^{1/4}}{(r^2 f)^{1/4}}K_{n}(k\xi(r)^{1/2})\sum^{j}_{s=0} \frac{A_s(\xi(r))}{k^{2s}}\nonumber\\
& -\frac{\xi(r)^{3/4}}{k(r^2 f)^{1/4}}K_{n+1}(k\xi(r)^{1/2})\sum^{j-1}_{s=0} \frac{B_s(\xi(r))}{k^{2s}}\nonumber\\
& +\tilde{\epsilon}_{2j+1,2}(k,\xi(r)),{}
\end{align}
\end{subequations}
where $\tilde{\epsilon}=\epsilon/ (\xi r^2 f)^{1/4}$.
It can be shown \cite{Olver} that for fixed $r$ and large $k$
\begin{equation*}
\tilde{\epsilon}_{2j+1,i} \approx O( {k^{-2j - 1}}),\qquad i=1,2
\end{equation*}
uniformly with respect to $r \in [r_0,r_b]$, provided $f$ does not vanish in $(r_0,r_b]$.
So we can see that the zeroth order extended Green-Liouville (EGL)  approximation ($j=0$) is an approximation to $O(l^{-1})$ for large $l$, the first $O(l^{-3})$ etc. Therefore for large $l$
\begin{subequations}
\begin{align}
C_{n l}p_{n l}(r)q_{n l}(r)-P_{0}(r)Q_{0}(r)= O({l^{-3}}),\\
\label{order}
C_{n l}p_{n l}(r)q_{n l}(r)-P_{1}(r)Q_{1}(r)= O({l^{-5}}).
\end{align}
\end{subequations}
Inspection of Eqs.~(\ref{num}) and (\ref{order}) shows that subtracting the EGL approximation to first order will render the sum over $l$ $O(1/l^4)$ making the sums converge rapidly. A comparison plot between the exact solution and its zeroth order EGL approximation $P_0Q_0$ can be found in Fig.~\ref{fig:Compar}

In our derivation of the EGL approximation in Sec.~\ref{GreenLiouvilleA}  we implicitly assumed that $x_0$ was the only singular point of  our differential equation in the region under consideration. However for a black hole space-time with multiple horizons, such as our  lukewarm black-hole configuration, the radial equation with have a second singular point, at $r_1$, say. (For the lukewarm black hole, we will take $r_0$ to be the event horizon of the black hole and  $r_1$ to be the cosmological horizon.) In this case, our original approximation fails to be uniform as we approach $r_1$ and   we need a second EGL approximation based on an expansion around $r_1$. The derivation proceeds along the same lines as before yielding the following approximations for $p_{n l}$ and $q_{n l}$, respectively:
\begin{subequations}
\begin{align}
\label{expanh}
\hat{P}_{j}(k,r)& =\frac{\hat{\xi}(r)^{1/4}}{(r^2 f)^{1/4}}K_{n}(\hat{k}\hat{\xi}(r)^{1/2})\sum^{j}_{s=0} \frac{\hat{A}_s(\hat{\xi}(r))}{\hat{k}^{2s}} \nonumber\\
&-\frac{\hat{\xi}(r)^{3/4}}{\hat{k}(r^2 f)^{1/4}}K_{n+1}(\hat{k}\hat{\xi}(r)^{1/2})\sum^{j-1}_{s=0} \frac{\hat{B}_s(\hat{\xi}(r))}{\hat{k}^{2s}}\nonumber\\
 &+\hat{\tilde{\epsilon}}_{2j+1,1}(\hat{k},\hat{\xi}(r));{}\\
\hat{Q}_{j}(k,r) &=\frac{\hat{\xi}(r)^{1/4}}{(r^2 f)^{1/4}}I_{n}(\hat{k}\hat{\xi}(r)^{1/2})\sum^{j}_{s=0} \frac{\hat{A}_s(\hat{\xi}(r))}{\hat{k}^{2s}} \nonumber\\
&-\frac{\hat{\xi}(r)^{3/4}}{\hat{k}(r^2 f)^{1/4}}I_{n+1}(\hat{k}\hat{\xi}(r)^{1/2})\sum^{j-1}_{s=0} \frac{\hat{B}_s(\hat{\xi}(r))}{\hat{k}^{2s}}\nonumber\\
 &+\hat{\tilde{\epsilon}}_{2j+1,2}(\hat{k},\hat{\xi}(r)),{}
 \end{align}
\end{subequations}
where
\begin{align*}
\hat{k}^2 &= l(l+1) +\left(m^2 +\left(\xi-\textstyle{\frac{1}{6}}\right)R(r_1)\right)r_1^2\nonumber\\
&\qquad +n^2\bigg(\frac{R(r_1)}{6} +2\kappa_1 r_1\bigg) +\frac{1}{3}(1-n^2)\nonumber\\
&\equiv l(l+1) + \hat{N}\nonumber\\
\hat{ \xi} &=\bigg(\int_{r}^{r_1} \frac{1}{\sqrt{r'^2 f}} dr'\bigg)^2.
\end{align*}
and where  $\hat{A}_s$ and $\hat{B}_s$ satisfy the recursion relations Eq.~(\ref{recur}) with $N$ replaced by $\hat{N}$ in the definition of $\psi$. The same large $l$ behavior applies to $\hat{\tilde{\epsilon}}_{1,2}$ as in the previous case.

We now have approximations $P$ and $Q$ to the functions 
$p_{n l}$ and $q_{\omega l}$ respectively that are uniform in $l$ over a closed region including the event horizon but not inclusive of the cosmological horizon and vice versa for $\hat{P}$ and $\hat{Q}$. Our approach, which will be discussed in more detail in Sec.~\ref{sec:FormalCalculation},  then is to subtract  and add on both approximations by using a switching function to alternate between them.

\section{Formal Calculation of $\langle\hat{\varphi}^2\rangle_{ren}$}
\label{sec:FormalCalculation}

We now turn to the details of the calculation of $\langle\hat{\varphi}^2\rangle_{ren}$ for  lukewarm black holes. We will first briefly review the key features of these spacetimes before describing the details of our calculations.
\subsection{Lukewarm Black holes}
\label{sec:FormalCalculationA}
Lukewarm black holes are a special class Riessner-N\"ordstrom-de Sitter spacetimes with line element given by Eq.~(\ref{le}) with
\begin{equation}
\label{metric}
f (r) = 1 -\frac{2M}{r} +\frac{Q^2}{r^2} - \frac{\Lambda r^2}{3},
\end{equation}
where $M$, $Q$ are the mass and charge of the black hole respectively and $\Lambda$ is the (positive) cosmological constant, with $Q=M$. 
The Ricci tensor is given by
 \begin{align}
 R_{\mu\nu} = \begin{cases} \left(\Lambda -\displaystyle{\frac{Q^2}{r^4}}\right)g_{\mu \nu}& \quad \mu,\nu=\tau, r\\  \left(\Lambda +\displaystyle{\frac{Q^2}{r^4}}\right)g_{\mu \nu}& \quad \mu,\nu=\theta, \phi \end{cases}
  \end{align}
giving a constant Ricci scalar $R=4\Lambda$.
For $4M<\sqrt{3/\Lambda}$ we have three distinct horizons, a black-hole event horizon at $r=r_h$, an inner Cauchy horizon at $r=r_-$, and a cosmological horizon at $r=r_c$, where
\begin{subequations}
\begin{align}
r_-&=\frac{1}{2}\sqrt{{3}/{\Lambda}}\left(-1 +\sqrt{1 +4M\sqrt{{\Lambda}/{3}}}  \right)\\
r_h&=\frac{1}{2}\sqrt{{3}/{\Lambda}}\left(1 -\sqrt{1 -4M\sqrt{ {\Lambda}/{3}}}  \right)\\
r_c&=\frac{1}{2}\sqrt{{3}/{\Lambda}}\left(1 +\sqrt{1 -4M\sqrt{ {\Lambda}/{3}}}  \right).
\end{align}
\end{subequations}
The 4th  root of $f$ is negative and hence nonphysical. The Penrose diagram for this spacetime can be found in~\cite{Mellor}.

The surface gravities of event and cosmological horizons coincide and are given by
\begin{equation*}
\kappa_{h}=\kappa_{c}=\sqrt{\Lambda/3}\sqrt{1 -4M\sqrt{ {\Lambda}/{3}}} .
\end{equation*}
 We shall confine attention to the region $r \in [r_h,r_c]$, which has a regular Euclidean section with topology $S^2 \times S^2$ \cite{Mellor}. We will
take our quantum field to be in a thermal state at the natural temperature $T=\kappa/2 \pi$ where we have dropped the subscript for notational compactness.

\subsection{ Computational Strategy}
\label{sec:FormalCalculationB}
It is known that the divergence on both the event and cosmological horizons for $\langle\hat{\varphi}^2\rangle_{numeric}$ is entirely contained in the $n=0$ mode  \cite{Winstanley:2007} , so the WKB approximation only encounters the problems outlined in Sec.\ref{sec:PreviousResultsB} for this mode. As the WKB mode is easier to work with than the EGL approximation, our computational strategy is  to use a combination of both WKB and EGL approximations. We make use of EGL for the $n=0$ mode and WKB for all other $n$ modes.  For $\langle\hat{T}_{\mu\nu}\rangle$ the corresponding strategy requires the use of the EGL approximation for  the $n=0$, 1 and 2 modes. 
\subsection{WKB approximation}
\label{sec:FormalCalculationC}
We adopt the WKB approach of Casals et.~al~\cite{Barry} which we will now briefly outline.
We begin by defining $\beta(r)$ by
\begin{equation*}
\beta(r) = C_{n l}p_{n l}(r)q_{\omega l}(r).
\end{equation*}
Then $\beta(r)$ satifies the nonlinear differential equation
\begin{equation}
\label{beqn}
r^2 f \frac{\mathrm{d}\ }{\mathrm{d}r}\bigg( r^2 f\frac{\mathrm{d} \beta^{1/2}}{\mathrm{d}r\ }\bigg) -(\eta^2 +\chi^2)\beta^{1/2}+\frac{1}{4\beta^{3/2} }=0,
\end{equation}
where
\begin{align*}
\chi(r) =\sqrt{n^2 \kappa^2 r^4 +\left(l+\textstyle{\frac{1}{2}}\right)^{\!2} r^2 f},\nonumber\\
\eta(r) =\textstyle{\frac{1}{4}} f r^2 +(m^2 +\xi R) f r^4.
\end{align*}
We are looking for a large $l$ and/or large $\omega$ approximation, so we seek to express $\beta(r)$ as an expansion in inverse powers of $\chi(r)$. To keep track of orders it is convenient to replace $\chi$ in Eq.~(\ref{beqn}) by $\chi/\epsilon$ where $\epsilon$ is an expansion parameter which we will finally set equal to $1$ at the end of the calculation. We then write
\begin{equation*}
\beta(r) =\epsilon \beta_0(r) +\epsilon^2 \beta_1 (r) +...
\end{equation*}
To balance Eq.~(\ref{beqn}) to leading order we require
\begin{equation*}
 \beta_0(r) =\frac{1}{2 \chi(r)}.
\end{equation*}
Inserting this into Eq.~(\ref{beqn}) for $\beta$ and solving formally order by order in $\epsilon$ we find a recursion relation for the $\beta_n(r)$. On doing so we obtain the following expansion
\begin{equation}
\label{wkb cof}
\beta_{n} (r) = \sum_{m=0}^{2n} \frac{A_{n,m}(r) n^{2m}}{\chi^{2n +2m +1}}
\end{equation}
where a recursion relation for the $A_{n,m}(r)$ can be obtained from rearrangement of the $\beta_n(r)$ recursion relations~\cite{Barry} .
It can be seen from Eq.~(\ref{wkb cof}) that, for large $\chi$ (or equivalently large $l$ or $n$ if $f\neq0$),
$
\beta_{n} \sim \chi^{-2n-1}
$.
So for fixed $r$ and large $l$ we have the same asymptotic behavior as the EGL approximation. Hence if we consider the double sum
\begin{equation}
\sum^{\infty}_{n=1}\sum_{l=0}^{\infty} (2l+1)[C_{n l}p_{n l}(r)q_{\omega l}(r) -\beta_0 (r) -\beta_1 (r)],
\end{equation}
then the summand is $O(l^{-4})$ for large $l$. In addition, we see that once the sum over $l$ is completed, the resulting summand is $O(n^{-3})$. Therefore in order to compute the mode sums in 
$\langle\hat{\varphi}^2\rangle_{numeric}$ given by Eq.~(\ref{num}) it is sufficient to subtract either the EGL or  WKB approximations to the first order.

\subsection {Incorporating the Approximations}
We now wish to incorporate these approximations into expressions for $\langle\hat{\varphi}^2\rangle_{numeric}$ and $\langle\hat{\varphi}^2\rangle_{analytic}$. We do this in the following manner, which will be justified later on. We write
\begin{align}
\label{num2}
&\langle\hat{\varphi}^2\rangle_{numeric}=\frac{\kappa}{4 \pi^2}\sum_{n=1 }^{\infty} \bigg\{\frac{n \kappa}{f} +\frac{1}{2n\kappa}\left(m^2 +\left(\xi-\textstyle{\frac{1}{6}}\right) R\right) \nonumber\\
&+\sum^{\infty}_{l=0}(2l+1) \left( C_{n l}p_{n l}(r)q_{n l}\left(r\right)
-\beta_0(r) -\beta_1(r)\right)\nonumber\\
&+\sum^{\infty}_{l=0}\left[(2l+1)\left( \beta_0(r) +\beta_1(r)\right)
-\frac{1}{r f^{1/2}}\right]\bigg\}\nonumber\\
&+\frac{\kappa}{8 \pi^2}\sum^{\infty}_{l=0}(2l+1) \left( C_{0 l}p_{0 l}(r)q_{0 l}(r)\right.\nonumber\\
&\left. -\mathcal{S}(r)(\gamma_0(r) +\gamma_1(r))-\hat{\mathcal{S}}(r)\left(\hat{\gamma}_0 (r) +\hat{\gamma}_1 (r) \right)\right)\nonumber\\
&+\frac{\kappa}{8 \pi^2}\sum^{\infty}_{l=0}(2l+1)\left(\mathcal{S}(r)\gamma_1 (r) +\hat{\mathcal{S}}(r)\hat{\gamma}_1 (r)\right) ,
\end{align}
\begin{align}
\label{anal2}
&\langle\hat{\varphi}^2\rangle_{analytic}=\frac{\kappa}{8 \pi^2}\mathcal{S}(r)\sum^{\infty}_{l=0}\left[(2l+1)\gamma_0(r)- \frac{1}{r f^\frac{1}{2}}\right]\nonumber\\
&+\frac{\kappa}{8 \pi^2}\hat{\mathcal{S}}(r)\sum^{\infty}_{l=0}\left[(2l+1)\hat{\gamma}_0(r)- \frac{1}{r f^{1/2}}\right]\nonumber\\
&- \frac{1}{8\pi^2}\left(m^2 +\left(\xi-\textstyle{\frac{1}{6}}\right)R\right)\ln\left(\frac{\mathrm{e}^{2\gamma}\mu^2f}{ 4 \kappa_0{}^2}\right)\nonumber\\ 
&\quad +\frac{m^2}{16\pi^2} -\frac{f'{}^2}{192\pi^2 f} 
  +\frac{f''}{96\pi^2 }+\frac{f'}{48\pi^2 r} +\frac{\kappa_0{}^2}{48 \pi^2 f},
\end{align}
here $\mathcal{S}(r)$ is a switching function, $\hat{\mathcal{S}}(r)= 1- \mathcal{S}(r)$ and $\hat{\gamma}$ denotes the  EGL approximation valid on the cosmological horizon. The form of $\mathcal{S}(r)$ used is irrelevant. The only condition on it is that  it has  a value of  $1$ on the event horizon and of $0$ on the cosmological horizon. In this case we found the form 
\begin{align*}
\mathcal{S}(r)=\tanh\bigg(\frac{r_c-r}{r-r_h}\bigg)^3,
\end{align*}
to be the most convenient. 

The analytical sums of WKB approximation over $l$ have already been computed in  \cite{Winstanley:2007} so  we will omit the details and just present the results
\begin{align}
&\sum^{\infty}_{l=0}\bigg[(2l+1)\left(\beta_0(r) +\beta_1(r)\right)-\frac{1}{r f^\frac{1}{2}}\bigg] =-\frac{\omega}{f}\nonumber\\
& -\frac{1}{2\omega}\left(m^2 +\left(\xi-\textstyle{\frac{1}{6}}\right)R\right) +J_0(\omega,r) +J_1(\omega,r), 
\end{align}
where $J_0(\omega,r)$ and $J_1(\omega,r)$ are integrals which can be found in~\cite{Winstanley:2007} and are easily evaluated numerically.

We now turn our attention the the sum
\begin{align}
\label{GLsum}
&\sum^{\infty}_{l=0}\left[(2l+1)\gamma_0(r)- \frac{1}{r f^\frac{1}{2}}\right]=\nonumber\\
&\sum^{\infty}_{l=0}\left[(2l+1)A(r)I_0(k\zeta)K_0(k \zeta)-\frac{1}{r f^{1/2}}\right],
\end{align}
where
\begin{equation}
\label{A}
 A(r)=\frac{\zeta}{r f^{1/2}}.
\end{equation}
and we have, for convenience, defined $\zeta=\xi^{1/2}$.
We compute this sum using the 
Watson-Sommerfield formula, valid for any function analytic in the right-hand half plane:
\begin{align*}
\sum_{l=0}^{\infty} F(l) &= \int_{0}^{\infty} F\bigg(\lambda- \frac{1}{2}\bigg)\mathrm{d}\lambda\nonumber\\
&-\mathcal{R} \bigg[ i  \int_{0}^{\infty}\frac{2}{1+e^{2 \pi \lambda}} F\bigg(i\lambda- \frac{1}{2}\bigg)\mathrm{d}\lambda\bigg].
\end{align*}
Applying this to Eq.~(\ref{GLsum}) we can express the sum as a sum of two integrals 
\begin{subequations}
\begin{align*}
&\mathcal{I}_1(r)=\int^{\infty}_{0} \left[2\lambda A(r)I_0(k_\lambda \zeta)K_0(k_\lambda \zeta)-\frac{1}{r f^{1/2}}\right]\mathrm{d}\lambda\\
&\mathcal{I}_2(r)= \mathcal{R}\bigg[\int^{\infty}_{0}\!\frac{ 4\lambda}{1 +e^{2 \pi \lambda}} A(r) I_0(k_{i \lambda} \zeta)K_0(k_{i \lambda }\zeta)\mathrm{d}\lambda\bigg]
\end{align*}
\end{subequations}
with
\begin{align}
\label{k}
k_\lambda^2=\lambda^2 +N +\textstyle{\frac{1}{12}},\quad\quad k_{i\lambda}^2=-\lambda^2 +N +\textstyle{\frac{1}{12}}.
\end{align}

\subsection*{Evaluation of $\mathcal{I}_1(r)$} 
Inspecting Eq.~(\ref{k}) we can see that
$
2 k \mathrm{d}k =2 \lambda \mathrm{d}\lambda
$
and so we can change $I_1$ to an integral over $k$ as follows
\begin{align}
\label{Ik}
\mathcal{I}_1(r) = \int_{k_0}^{\infty} \bigg[2 k A(r)I_0(k \zeta)K_0(k \zeta)-\frac{\mathrm{d}\lambda}{\mathrm{d} k}\frac{1}{r f^{1/2}}\bigg] \mathrm{d}k.\nonumber\\
\end{align}
To compute this integral we first introduce a large $k$ cutoff $k_l$ and then we take the limit $k_l \to \infty$ after the integration is performed. We have \cite{Watson} 
\begin{equation*}
\int 2 k I_0( k v)K_0( kv) \mathrm{d}k =k^2[  I_0( k v)K_0( kv) + I_1( k v)K_1( kv)],
\end{equation*}
and so Eq.~(\ref{Ik}) evaluates to
\begin{align}
\label{lkI}
\mathcal{I}_1(r)&= \lim_{k_l\to \infty}\bigg[A(r) k^2[I_0(k \zeta)K_0(k \zeta)\nonumber\\
&\qquad\qquad+I_1(k \zeta)K_1(k \zeta)]-\frac{\lambda(k)}{r f^{1/2}}\bigg]_{k_0}^{k_l}.
\end{align}
For large $z$, we have the expansion \cite{Watson}
\begin{equation*}
I_n(z)K_n(z) = \frac{1}{2 z}\bigg[1 - \frac{4n^2 -1}{8z^2} +O(z^{-4})\bigg].
\end{equation*}
Using this together with Eqs.~(\ref{A}) and (\ref{k}) we see that the upper limit contribution in Eq.~(\ref{lkI}) is $O(k_l^ {-1})$ 
and so we arrive at the result
\begin{align}
\mathcal{I}_1(r)=&- A(r)k_0^2[I_0(k_0 \zeta)K_0(k_0 \zeta)+I_1(k_0 \zeta)K_1(k_0 \zeta)].
\end{align}
with $k_0^2 = N +1/12$.

\subsection*{Evaluation of $\mathcal{I}_2(r)$} 

\begin{figure}[htb!]
\centering
\includegraphics[width=8cm]{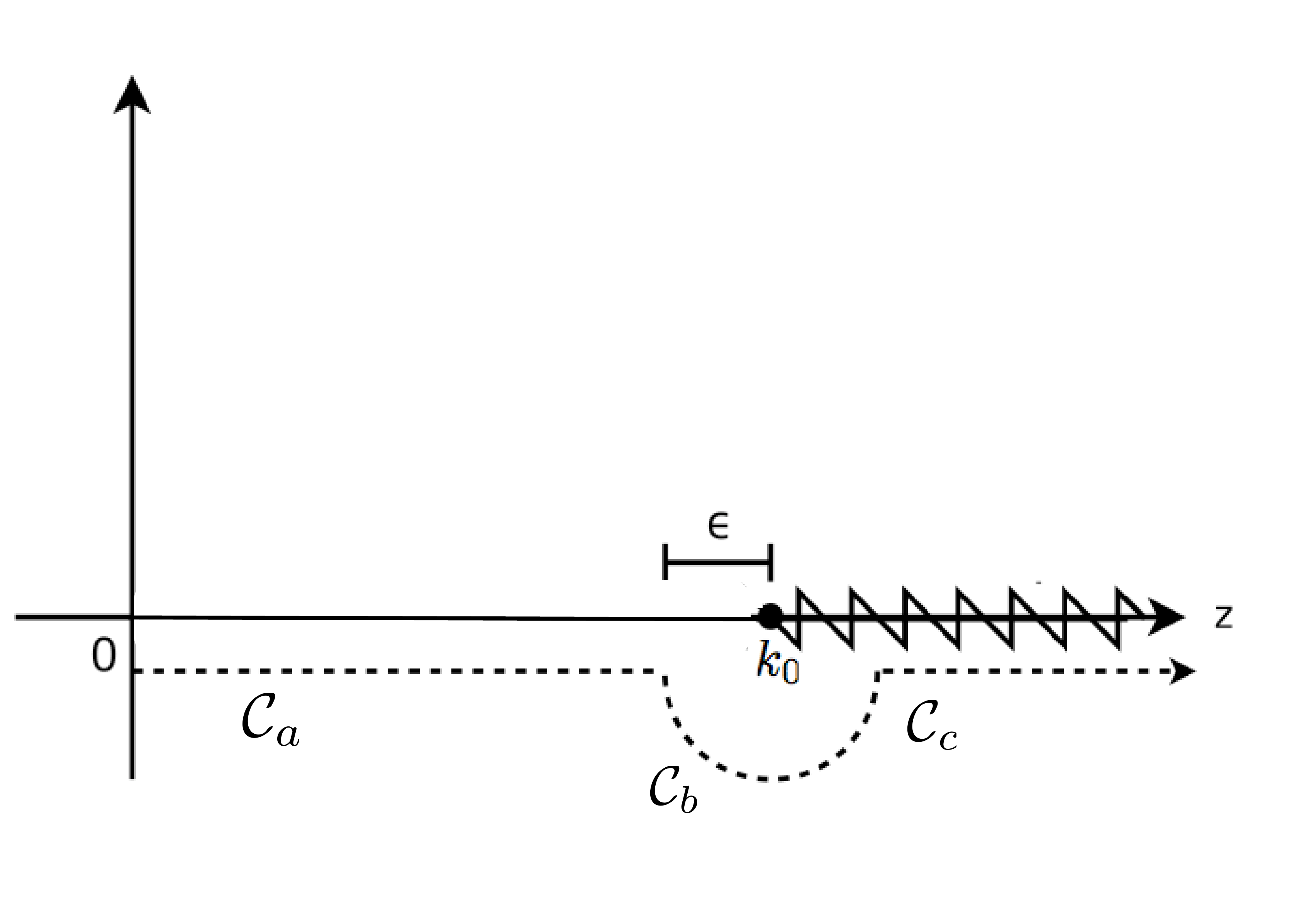}\caption{\label{fig:ITwocontour} Contour of integration for $\mathcal{I}_2$.}
\end{figure}

We now examine the second integral. First we note that $k_{i\lambda}=(k_0{}^2-\lambda^2)^{1/2}$ goes to zero at $\lambda= | k_0| $, hence we must introduce cuts in the complex $\lambda$-plane which we take to run from   $\lambda =  -k_0$ to $-\infty$ and from $\lambda =  k_0$ to $\infty$. So we consider the contour of integration shown in Fig.~\ref{fig:ITwocontour}.
We define $\mathcal{I}_{2a}$, $\mathcal{I}_{2b}$ and $\mathcal{I}_{2c}$ as the contributions from $\mathcal{C}_{a}$, $\mathcal{C}_{b}$ and $\mathcal{C}_{c}$ respectively.

Considering $\mathcal{I}_{2a}$ first, we allow our variable to run from 0 to $k_0-\epsilon$, also as the expression is real we can drop the $\mathcal{R}$ symbol. Hence the integral becomes
simply
\begin{equation*}
\mathcal{I}_{2a}=\int_{0}^{k_0-\epsilon}  \frac{4\lambda d\lambda}{1+e^{2 \pi \lambda}} \left(A(r)I_0(k_{i \lambda} \zeta)K_0(k_{i \lambda }\zeta)\right)
\end{equation*}
$I_{2a}$ contains a divergence as we approach the event horizon, i.e as $\zeta \to 0$. We choose to take this out of the integral explicitly as follows
\begin{align}
\label{I2a}
I_{2a} &=\int_{0}^{k_0-\epsilon}  \frac{4\lambda d\lambda}{1+e^{2 \pi \lambda}} \bigl(A(r)I_0(k_{i \lambda} \zeta)K_0(k_{i \lambda }\zeta)+A_h\ln{\zeta}\bigr)\nonumber\\
&\qquad- A_h \ln{\zeta} \int_{0}^{k_0-\epsilon}  \frac{4\lambda d\lambda}{1+e^{2 \pi \lambda}} 
\end{align}
with $A_h\equiv A(r_h)= {1}/{(\kappa r_h^2)}$.
Details of the expansions used above can be found in Appendix A.

$I_{2b}$ can easily be shown to be $O(\epsilon\ln(\epsilon))$ and so does not contribute.

Lastly we consider the integral $I_{2c}$. We allow our variable $\lambda$ to run from $k_0+\epsilon$ to $\infty$. In this region $k_{i\lambda}$ is purely imaginary, so we define $\hat{k}=ik_{i\lambda}$ and use the following identity  \cite{Watson} :
\begin{equation*}
I_0(ix)K_0(ix)=-\frac{\pi}{2}\left(J_0(x)Y_0(x) +  iJ_0(x)^2\right)
\end{equation*}
giving
\begin{equation*}
I_{2c}=-\int^{\infty}_{k_0}  \frac{2\pi\lambda d\lambda}{1+e^{2 \pi \lambda}} A(r) J_0(\hat{k}\zeta)Y_0(\hat{k}\zeta)
\end{equation*}
with $\hat{k}=\sqrt{\lambda^2-k_0^2}$. Proceeding along the lines of the $I_{2a}$ calculation yields the following expression for $I_2$
\begin{align*}
 I_{2}(r) &=-\frac{A_h}{12} \ln\zeta\nonumber\\
&+ \int_{0}^{k_0}\frac{4\lambda d\lambda}{1+e^{2 \pi \lambda}} \bigl(A(r)I_0(k_{i \lambda} \zeta)K_0(k_{i \lambda }\zeta)+ A_h\ln{\zeta}\bigr)\nonumber\\
& + \int^{\infty}_{k_0}  \frac{4\lambda d\lambda}{1+e^{2 \pi \lambda}} \left(-\frac{ \pi}{2}A(r) J_0(\hat{k}\zeta)Y_0(\hat{k}\zeta)
 +A_h\ln{\zeta}\right) .
\end{align*}

Finally we have the result 
\begin{align}
\label{GLsum2}
&  \Gamma(r) \equiv \sum^{\infty}_{l=0}\left[(2l+1)\gamma_0(r)- \frac{1}{r f^\frac{1}{2}}\right]=-\frac{A_h}{12} \ln\zeta\nonumber\\
&+\int_{0}^{k_0}  \frac{4\lambda d\lambda}{1+e^{2 \pi \lambda}} \bigg(A(r)I_0(k_{i \lambda} \zeta)K_0(k_{i \lambda }\zeta)
+A_h \ln{\zeta}\bigg)\nonumber\\
& + \int^{\infty}_{k_0}  \frac{4\lambda d\lambda}{1+e^{2 \pi \lambda}} \bigg(-\frac{A(r) \pi}{2} J_0(\hat{k}\zeta)Y_0(\hat{k}\zeta)+ A_h\ln\zeta\bigg)\nonumber\\
  &-A(r) k_0^2\{I_0(k_0 \zeta)K_0(k_0 \zeta+I_1(k_0 \zeta)K_1(k_0 \zeta)\} ,
\end{align}
and similarly for $\hat{\gamma}_0(r)$ with $\zeta$ and $N$ replaced by $\hat{\zeta}$ and $\hat{N}$,respectively.
It can now easily be shown, using the expansions in Appendix \ref{Ap:NearHorizon}, that near the event horizon
\begin{equation*}
\frac{\kappa}{8 \pi^2} \Gamma(r) = \frac{1}{16 \pi^2} \left(m^2 +\left(\xi-\textstyle{\frac{1}{6}}\right)R\right) \ln(r-r_h) + O(1)
\end{equation*}
which ensures $\langle\hat{\varphi}^2\rangle_{analytic}$ finite on the event horizon. The same analysis applied to $\hat{\Gamma}(r)$ shows that $\langle\hat{\varphi}^2\rangle_{analytic}$ is also finite on the cosmological horizon. From this we can also deduce that the contribution of the sums $\sum^{\infty}_{l=0}(2l+1)\gamma_1(r)$ and $\sum^{\infty}_{l=0}(2l+1)\hat{\gamma}_1(r)$  to $\langle\hat{\varphi}^2\rangle_{numeric}$ is finite on both the event and cosmological horizons.

At this stage it is helpful to bring together the results of this section so far giving
\begin{align}
\label{numf}
&\langle\hat{\varphi}^2\rangle_{numeric}=\frac{\kappa}{4 \pi^2}\sum_{n=1 }^{\infty} \bigg\{
\sum^{\infty}_{l=0}\bigg[(2l+1) C_{n l}p_{n l}(r)q_{\omega l}(r)\nonumber\\
&\qquad-\beta_0(r) -\beta_1(r)\bigg]+Jo(\omega,r ) + J_1 (\omega,r)\bigg\}\nonumber\\
&+\frac{\kappa}{8 \pi^2}\sum^{\infty}_{l=0}\bigg[(2l+1) C_{0 l}p_{0 l}(r)q_{0 l}(r)\nonumber\\
&\qquad-\mathcal{S}(r)(\gamma_0(r) +\gamma_1(r))
-\hat{\mathcal{S}}(r)(\hat{\gamma}_0(r)+\hat{\gamma}_1(r))\bigg]\nonumber\\
&+\frac{\kappa}{8 \pi^2}\sum^{\infty}_{l=0}(2l+1)[\mathcal{S}(r)\gamma_1(r) +\hat{\mathcal{S}}(r)\hat{\gamma}_1(r)],
\end{align}
\begin{align}
\label{analf}
&\langle\hat{\varphi}^2\rangle_{analytic}=\frac{\kappa}{8 \pi^2}\bigl(\mathcal{S}(r)\Gamma(r)+\hat{\mathcal{S}}(r))\hat{\Gamma}(r)\bigr)\nonumber\\
&\quad- \frac{1}{16\pi^2}\left(m^2 +\left(\xi-\textstyle{\frac{1}{6}}\right)R\right)\ln\left(\frac{\mathrm{e}^{2\gamma}\mu^2f}{ 4 \kappa_0{}^2}\right)\nonumber\\ 
&+\frac{m^2}{16\pi^2} -\frac{f'{}^2}{192\pi^2 f} 
  +\frac{f''}{96\pi^2 }+\frac{f'}{48\pi^2 r} +\frac{\kappa_0{}^2}{48 \pi^2 f}. 
\end{align}
Both of which are now manifestly finite on the horizons. Clearly $\langle\hat{\varphi}^2\rangle_{analytic}$ contains expressions that can only be evaluated numerically, so the label ``analytic" here is used for the purpose of comparison  of our results with those of existing methods \cite{Winstanley:2007,Anderson:1994hg}.

The sums in $\langle\hat{\varphi}^2\rangle_{numeric}$ are now amenable to numerical computation, which will be discussed in the next section.
\section{Numerical calculations}
\label{sec:NumericalCalculations}

To perform the mode-sums contained in $\langle\hat{\varphi}^2\rangle_{numeric}$ we first need to find the modes themselves. We do this by numerical integration of the radial equation (\ref{mode}). We perform all of our numerical calculations using \textsl{Mathematica}~\cite{Mathematica}. Previous calculations  \cite{Anderson:1990jh, Anderson:1994hg, JensenOttewill:89,JensenOttewill:95,Howard:1984qp}  relied on programming languages such as C, C++, or Fortran. The use of a higher level application  greatly speeded up the formulation of the code required to calculate the quantities required to construct $\langle\hat{\varphi}^2\rangle_{ren}$.

Turning our attention to calculating the modes, the radial equation (\ref{mode}) has regular singular points at $r=r_h$ and $r=r_c$,  we apply the standard Frobienus method, by writing
\begin{equation}
S_{\omega l} =\sum^{\infty}_{i=0} a_i (r-r_0)^{i + \alpha},
\end{equation}
to obtain the indicial equation $\alpha = \pm n/2$.
$S$ is either $p$ or $q$, and  $r_0$ denotes the singular point under consideration.
We chose our modes so that $p_{n l}$ is  the solution regular on the event horizon, divergent on the cosmological horizon and vice versa for  $q_{\omega l}$. The standard procedure, when one is dealing with an asymptotically flat space time, is to obtain $p_{n l}$  by direct integration of Eq.~(\ref{mode}) with initial conditions given by the value of the series solution
\begin{equation}
\label{series p}
p_{n l} =\sum^{\infty}_{i=0} a_i (r_s-r_h)^{i + n/2},
\end{equation}
at a point $r=r_s$ near $r=r_h$ and then using the Wronskian condition (\ref{Wron}) to obtain $q_{\omega l}$. However in the case under consideration we have a finite outer boundary ($r=r_c$) hence we can also solve directly for $q_{\omega l}$ by integrating Eq.~(\ref{mode}) from a point outside the cosmological horizon $r=r_s'$, with initial condition
\begin{equation}
\label{series q}
q_{\omega l} =\sum^{\infty}_{i=0} b_i (r_c-r_s')^{i +n/2}.
\end{equation}
The $b_i$ satisfy a seven-term recursion relation, given in Appendix \ref{Ap:RecursionRelations}.
In order for our numerical integration to retain a high degree of accuracy, we found it advantageous to take our initial point away from the vicinity of the horizon. We found that a point about $(r_c -r_h)/10$ from each horizon to be most convenient.
 It is known that a series solution about a regular singular point of a differential equation has a radius of convergence of a least the distance to the next regular singular point \cite{Leaver}. Since the nearest regular singular point to $r=r_c$ is the event horizon, we see that the series (\ref{series q}) is valid $\forall r \in (r_h,r_c]$. This is not the case for the series (\ref{series p}) however, as the nearest regular singular point is given by the Cauchy horizon $r=r_i$ and since $|r_h-r_i| << |r_c-r_h|/10$ this poses a problem. To get around this issue we define a new Jaffe-like series solution \cite{Leaver} for $p_{n l}$. We define a new variable
 \begin{equation*}
u =\frac{r-r_h}{r-ri},
\end{equation*}
which pushes the Cauchy horizon out to infinity and ensures that the nearest regular singular point is $r=r_c$.  We then have a new series solution for $p_{n l}$ 
\begin{equation}
p_{n l} =(r_h -r_i)^{n/2}(1-u)^6\sum^{\infty}_{j=0} c_j u^{j + n/2},
\end{equation}
which is now valid $\forall r \in [r_h,r_c)$. The coefficients $c_j$ are determined by a seven-term recursion relation, given in Appendix~\ref{Ap:RecursionRelations}.

We are now in a position to calculate the mode sums in (\ref{num2}). 
The rapid convergence of the modes sums in (\ref{num2}) rely on the cancelation of very large numbers. Therefore it is neccesary to calculate the mode functions with great accuracy. To ensure this we set our code to a precision of $24$ digits. 
As was predicted in Secs.~\ref{sec:GreenLiouvilleB} and \ref{sec:FormalCalculationB} the sums over $l$ and $\omega$, are $O(l^{-4})$ and $O(\omega^{-3})$, respectively.

Finally we turn to the sum 

$\sum^{\infty}_{l=0}(2l+1)[\mathcal{S}(r)\gamma_1+\hat{\mathcal{S}}(r)\hat{\gamma}_1]$. Since we know \textit{a priori} that this sum does not contribute to the divergence on either horizon and considering the technical difficulties which would be involved in computing this sum analytically, we  instead simply did it
numerically. The NSum algorithm in \textsl{Mathematica} computes the sum analytically up to a certain 
user-definable cut-off value of $l$ and then uses Richardson extrapolation to calculate the numerical tail of the sum. With the optimum choice of the cutoff value, the resulting sum had an error $\approx .002 \%$, more than sufficient for our purposes.

\section{$\langle\hat{\varphi}^2\rangle_{ren}$ on the Black-Hole Horizons.}
\label{sec:HorizonValue}
In this section, we introduce a new method of explicitly calculating a value for $\langle\hat{\varphi}^2\rangle_{ren}$ on both the event and cosmological horizon. We will perform the calculation on the event horizon, the method trivially  generalizes to the cosmological horizon.  We calculate the terms corresponding to  $\langle\hat{\varphi}^2\rangle_{numeric}$ and $\langle\hat{\varphi}^2\rangle_{analytic}$ for radial separation in turn.

\subsection{$\langle\hat{\varphi}^2\rangle_{numeric}$}
\label{sec:HorizonValueA}
Up to this point we have only considered the case when the points are separated in time. The final answer for $\langle\hat{\varphi}^2\rangle_{ren}$ is, however, independent of the choice of point separation. From Eq.~(\ref{series p}) if we let $r=r_h$ we see that $p_{\omega,l}$ vanishes for $n>0$ and so has a value $a_0=1/\sqrt{r_h^2\kappa}$ on the horizon. Hence separating in the radial direction and placing one point on the horizon yields the result
\begin{align}
\label{numr}
&\langle\hat{\varphi}^2\rangle_{numeric} =\frac{a_0 \kappa}{8 \pi^2}\sum^{\infty}_{l=0}(2l+1)\bigg[ q_l(r) -A^{1/2}(r) K_0(k \zeta)\bigg],\nonumber\\
\end{align}
where we recall from Eq.~(\ref{A}) that 
$A(r)=\zeta/\sqrt{r^2 f}$.

To proceed any further we need an expression for  $q_{0l}(r)$. We will show that the zeroth order EGL approximation captures the required local behavior  of $q_{0l}(r)$ needed to calculate $\langle\hat{\varphi}^2\rangle_{ren}$ on the event horizon. It does not, indeed cannot, contain any global information. We find an expression for this global contribution, which needs to be evaluated numerically for the space-time under consideration. In the following we have been inspired by the method due to Candelas  \cite{Candelas:1980zt}  to express $q_{0l}(r)$ in terms of a local analytical term  plus a numerical global term which can be extended to the case of a lukewarm black hole, however we present a variation of that method, which provides a more natural way to extend to higher orders. 

We begin by noting that $q_{0l}(r)$ can be expanded in a series solution about the event horizon 
as~\cite{Burkill}
\begin{align}
\label{eq:Burk}
q_{0l}(r) &=- \frac{1}{2\sqrt{\kappa r_h^2}} \left(1  +\frac{V_h}{2(\kappa r_h^2)} \right) \ln(\epsilon) + \alpha_l p_{0l}(r)\nonumber\\
&+\frac{1}{2}\left[\left(\frac{f''_h}{2f'_h}+\frac{2}{r_h}+\frac{V_h}{(\kappa r_h^2)} \right)\epsilon \right]+O(\epsilon^2 \ln(\epsilon))
\end{align}
where  $V_h=l(l+1)+r_h^{2}(m^2 +\xi R)$ and  $\alpha_l$ depends only on $l$, $\epsilon=r-r_h$, $f'_h \equiv f'(r_h) $ and $f''_h \equiv f''(r_h) $.

Next we note that the singular zeroth order EGL approximation, $Q_{0}(r)\equiv{A}(r)^{1/2} K_0(k \zeta)$ satisfies
\begin{align}
\label{eq:GLeqn}
&\bigg[\frac{\mathrm{d}}{\mathrm{d}r}\bigg(r^{2}f\frac{\mathrm{d}}{\mathrm{d}r}\bigg)-\tilde{V}_l(r)\bigg]Q_0(r)= 0,
\end{align}
with
\begin{align*}
\tilde{V}_l(r)&=k^2 -\frac{1}{4\zeta^2} -\frac{f}{4} +\frac{r^2f'^2}{16f}- \frac{r(3f' +r f')}{4}.
\end{align*}

Applying the standard Frobenius theory~\cite{Burkill} to Eq.~(\ref{eq:GLeqn}) we may find a series solution to $Q_{0}(r)$ about the event horizon. This series is in agreement with Eq.~(\ref{eq:Burk}) up to $O(\epsilon^2 \ln(\epsilon))$  up to a possible difference in the $\alpha_{l}$ term. Hence we may write 
\begin{equation}
\label{qexpan}
q_{0l}(r)=Q_{0}(r) +\beta_l p_{0l}(r) + R_l(r),
\end{equation} 
where $R_l(r)$ denotes the remainder terms and near the event horizon is $O(\epsilon^2\ln(\epsilon))$ as $\epsilon \to 0$. (Comparing our method with that of Candelas~\cite{Candelas:1980zt} we see that $R_l(r)$ is equivalent to the $\tilde{q}$ term.) The constants $\beta_l$ are determined by the requirement that $q_{0l}(r)$ is regular on the outer boundary and are calculated in Appendix~\ref{Ap:Beta}.

It can also be shown that for $l\geq(r-r_h)^{-1/2}$ the contribution of the terms $R_l(r) +\beta_l p_{0l}(r)$ cuts off exponentially in $l$. Therefore we have an expression for the Green's function, valid when one of the points is on the horizon
\begin{equation}
\label{Eq:GreenHorizon}
G_E(x,x')= \frac{a_0\kappa}{8 \pi^2}\sum_{l=0}^{\infty}(2l+1)\left(Q_{0}(r) +a_0\beta_l \right)+O(\epsilon \ln(\epsilon)).
\end{equation}
The convergence of the sum over $l$ of $(2l+1)\beta_l$ presents no problem as it can be seen numerically that the $\beta_l$ constants behave like $l^{-4}$ for large $l$.

Inspection of Eq.~(\ref{Eq:GreenHorizon}) leads us to conclude that while the expansion ($\ref{qexpan}$) is of sufficient order to facilate the calculation of $\langle\hat{\varphi}^2\rangle_{ren}$, a higher-order approximation is required to capture the behavior needed for the corresponding calculation of $\langle\hat{T}_{\mu \nu}\rangle_{ren}$ as it will involve derivatives of the Green's function.

Finally inserting the expression (\ref{qexpan}) into Eq.~(\ref{numr}) we find that on the event horizon
\begin{equation*}
\langle\hat{\varphi}^2\rangle_{numeric}=\frac{1}{8 \pi^2 r_h^2}\sum^{\infty}_{l=0}(2l+1)\beta_l.
\end{equation*}

\subsection{$\langle\hat{\varphi}^2\rangle_{analytic}$}
\label{sec:HorizonValueB}
For radial point splitting, with $\sigma=2 s^2$, where $s$ is the radial geodesic seapration
\begin{align}
&\langle\hat{\varphi}^2\rangle_{analytic}=\frac{a_0 \kappa}{8 \pi^2}\sum^{\infty}_{l=0}(2l+1){A}^{1/2}(r) K_0(k \zeta)\nonumber\\
&-\frac{1}{4 \pi^2 s^2} -\frac{1}{16 \pi^2}(m^2 +\left(\xi-\textstyle{\frac{1}{6}}\right)R)\ln\bigg(\frac{\mathrm{e}^{2\gamma}\mu^2 | \sigma|}{2}\bigg)\nonumber\\
 &+\frac{m^2}{16\pi^2} -\frac{1}{96\pi^2}R^{\alpha\beta}{s_{\alpha}s_{\beta}}
\end{align}

A great advantage of radial point splitting is that it allows one to integrate the line element directly for $s$. Since $t=t'$, $\theta=\theta'$, and $\hat{\varphi}=\hat{\varphi}'$ the line element becomes
$\mathrm{d}s^2 ={\mathrm{d}r^2}/{f}$
and hence
\begin{equation}
\label{int}
s=\int_{r_h}^{r} \frac{1}{\sqrt{f(r')}} \mathrm{d}r'.
\end{equation}

For a Ricci-flat space time, one can perform this integral exactly and hence obtain an expression for $s$ everywhere without recourse to approximation. For a general non-Ricci-flat space time, this is not the case, but we can expand the integrand in the
near horizon limit.  Integrating Eq.~(\ref{int}) near the horizon we find (see Appendix~\ref{Ap:NearHorizon}) 
\begin{align}
\label{analr}
&\langle\hat{\varphi}^2\rangle_{analytic}=\frac{a_0\kappa}{8 \pi^2}\sum^{\infty}_{l=0}(2l+1)A^{1/2}(r) K_0(k \zeta)-\frac{f'_h}{16 \pi^2 \epsilon}  \nonumber\\
&-\frac{1}{16 \pi^2}(m^2 +\left(\xi-\textstyle{\frac{1}{6}}\right)R)\ln\bigg(\frac{\mathrm{e}^{2\gamma}\mu^2 \epsilon}{f'_h}\bigg) \nonumber\\
&+\frac{m^2}{16\pi^2}+\frac{f'_h}{48\pi^2 r_h}+O(\epsilon)
 \end{align}
  with $\epsilon = r-r_h$ and $f'(r_h)\equiv f'_h$.
To evaluate this quantity  we first express the $\epsilon \to 0$ divergences in terms of infinite sums over $l$, then we bring them inside the existing sum over $l$ and finally we calculate the resulting convergent sum.
First, we transform variable from $r$ to $\eta$ using
\begin{equation*}
\eta=\cosh\bigg[\int_{r_h}^{r} \frac{\mathrm{d}r'}{\sqrt{r'^2 f(r')}}\bigg],
\end{equation*}
expanding this near the horizon and then reexpressing $\langle \hat{\varphi}^2 \rangle_{div}$ in terms of $\eta$ gives
\begin{align}
\label{e}
&\langle \hat{\varphi}^2 \rangle_{div}= \frac{1}{8 \pi^2 r_h^2 (\eta-1)} \nonumber\\
&-\frac{1}{16 \pi^2}(m^2 +\left(\xi-\textstyle{\frac{1}{6}}\right)R)
\ln\bigg(\frac{\mathrm{e}^{2\gamma}\mu^2 r_h^2  (\eta-1)}{2}\bigg) \nonumber\\
 &+ \frac{ R }{96 \pi^2 }
-\frac{f'_h}{48\pi^2 r_h} -\frac{m^2}{16\pi^2}+O(\eta -1).
\end{align}
We have \cite{gradriz}
\begin{align*}
\sum^{\infty}_{l=0} (2l+1) P_l(1)Q_l(\eta)=\frac{1}{\eta-1}\quad \quad\quad\quad\quad  &\eta>1,\\
\sum^{\infty}_{l=0}\frac{1}{(l+1) x^{l+1}} =\ln\bigg(\frac{x}{x-1}\bigg)\quad \quad\quad\quad\quad &x>1.
\end{align*}

We can reexpress the latter, using the transformation $x\to\sqrt{\eta-1}+1$, as
\begin{align*}
&\sum^{\infty}_{l=0}\frac{1}{(l+1) (\sqrt{\eta-1}+1)^{l+1}} =\ln\bigg(\frac{\sqrt{\eta-1}+1}{\sqrt{\eta-1}}\bigg)\nonumber\\
&=-\frac{1}{2}\ln(\eta-1) +O((\eta-1)^{1/2}).
\end{align*}
We incorporate these divergent sums into Eq.~(\ref{analr}) to give, on the horizon,
\begin{align}
\label{nsum}
&\langle \hat{\varphi}^2 \rangle_{analytic}=\lim_{\eta \to 1}\bigg\{\sum^{\infty}_{l=0}\frac{1}{8\pi^2 r_h^2}\bigg[ \frac{N}{(l+1) (\sqrt{\eta-1}+1)^{l+1}}\nonumber\\
&+(2l+1)\left(\frac{A^{1/2}(\eta)}{a_0}K_0(k \cosh^{-1}(\eta))
 -Q_l(\eta)\right) \bigg]\bigg\}+ C_h
\end{align}
with 
\begin{align*}
C_h&= \frac{ R }{96 \pi^2 }-\frac{f'_h}{48\pi^2 r_h} -\frac{m^2}{16\pi^2}\nonumber\\
 & +\frac{1}{16 \pi^2}(m^2 +\left(\xi-\textstyle{\frac{1}{6}}\right)R)
 \ln\bigg(\frac{\mathrm{e}^{2\gamma}u^2 r_h^2 }{2}\bigg).
\end{align*}
We now in turn write $\eta$ in terms of $\epsilon$ and denote the resulting summand as $F(l,\epsilon)$. Eq.~(\ref{nsum}) then becomes
\begin{equation*}
\lim_{\epsilon \to 0}\sum^{\infty}_{l=0}F(l,\epsilon) +C_h .
\end{equation*}
It is now appropriate to expand $F(l,\epsilon)$ as
\begin{equation}
\label{expane}
F(l,\epsilon)=F_0(l) +F_1 (x_l)\sqrt{\epsilon} +\Delta F(x_l,\epsilon),
\end{equation}
where
\begin{align}
\label{defF0}
F_0(l)&=\lim_{\epsilon \to 0} F(l,\epsilon)\nonumber\\
&=(2l+1)[\psi(l+1)-\ln(k)]+\frac{N}{l+1},\\
\label{defF1}
F_1(x_l)&=\lim_{\epsilon \to 0}\frac{1}{\sqrt{\epsilon}} F\bigg(\frac{x}{\sqrt{\epsilon}},\epsilon\bigg)\nonumber\\
&=x_l\bigg(\frac{\beta^2}{6 \alpha^2} K_0( \alpha x_l)-\frac{\alpha N }{ x_l} K_1( \alpha x_l) +\frac{N}{x_l^2}e^{-x \frac{\alpha} {\sqrt{2}}}\bigg)\\
\Delta F(x,\epsilon)&\equiv F(l,\epsilon) -F_0(l) - F_1(x_l)\sqrt{\epsilon},
\end{align}
with
\begin{align}
x_l=(l+1/2)\sqrt{\epsilon}; \quad 
\alpha=\sqrt{\frac{2}{\kappa r_h^2}}; \quad \beta= \sqrt{\frac{2 R}{\kappa^2 r_h^2}},
\end{align}
and $\psi$ is the polygamma function.
Using the definition of a Riemann integral we can write:
\begin{equation*}
\int_{0}^{\infty}  F_1(x) dx =\lim_{x_{l+1} \to x_l} \sum^{\infty}_{l=0} F_1(x_l) (x_{l+1} -x_l)
\end{equation*}
where $x_k$ is arbritary point $\in [x_{l+1}, x_l]$.
From the definition of $x$ we see that
\begin{equation*}
\int_{0}^{\infty}  F_1(x)  dx =\lim_{\epsilon \to 0} \sum^{\infty}_{l=0} F_1(x_l)  \sqrt{\epsilon}
\end{equation*}
since $x_k$ is arbitrary.
Returning to Eq.~(\ref{expane}) we now have
\begin{eqnarray}
\label{expan2}
&\lim\limits_{\epsilon \to 0}\sum^{\infty}_{l=0}F(l,\epsilon)=\sum^{\infty}_{l=0}F_0(l)+ \int_{0}^{\infty}  F_1(x) dx
\nonumber\\
 &+ \lim\limits_{\epsilon \to 0}\sum^{\infty}_{l=0} \Delta F(x_l,\epsilon).
\end{eqnarray}

Since $F_0(l) + F_1(x)\sqrt{\epsilon} $ are the first two terms in the Taylor series in $\epsilon$ of $F(l,\epsilon)$ we see that $\Delta F(x,\epsilon)$ must be of order $\epsilon$ or higher, in other words
\begin{equation*}
\Delta F= \epsilon G(x_l,\epsilon)
\end{equation*}
 for some $G$ which has an $\epsilon$ dependence of $O(1)$ or higher.
Applying the Riemann sum argument here gives:
\begin{equation*}
\sum^{\infty}_{l=0} \Delta F(x,\epsilon) =\sqrt{\epsilon} \int_{0}^{\infty}  G(x,\epsilon) dx.
\end{equation*}

$F(x,\epsilon)$, $F_0(l)$, and  $F_1(x)$ are all clearly integrable in $x$ and so $G(x,\epsilon)$ is, by construction,  also integrable. Hence we can conclude that
\begin{equation}
\label{expanef}
\lim_{\epsilon \to 0}\sum^{\infty}_{l=0}F(l,\epsilon)=\sum^{\infty}_{l=0}F_0(l)+ \int_{0}^{\infty}  F_1(x) dx.
\end{equation}
We have confirmed this result by a direct numerical calculation of $\sum^{\infty}_{l=0}F(l,\epsilon)$ as $\epsilon \to 0$.

The right-hand side of Eq.~(\ref{expanef}) can be calculated analytically( see Appendix~\ref{Ap:Analytic} for details) to give
\begin{align}
\label{panalh}
&\langle \hat{\varphi}^2 \rangle_{analytic}=\frac{1}{8\pi^2 r_h^2}\bigg[\frac{d}{dx} \zeta\left(x,\textstyle{\frac{1}{2}} + i\delta\right) +\frac{1}{12} +\frac{\beta^2}{6\alpha^4}\nonumber\\
&+\frac{d}{dx} \zeta\left(x,\textstyle{\frac{1}{2}} - i\delta\right)\bigg|_{x=-1} -i \delta\ln\left(\frac{\Gamma\left(x,\textstyle{\frac{1}{2}} + i\delta\right)}{\Gamma\left(x,\textstyle{\frac{1}{2}} - i\delta\right)}\right)\nonumber\\
&+ N\left(1  -\ln\left(\mu r_h \right)\right)  \bigg] -\frac{ R }{96 \pi^2 }+\frac{f'_h}{48\pi^2 r_h} +\frac{m^2}{16\pi^2},
\end{align}
on the event horizon. Here $\zeta(x,a)$ is the generalized Riemann zeta function, $\Gamma$ is the Euler gamma function and $\delta =\sqrt{N +1/12}$.

This method is trivially extended to the cosmological horizon, yielding a similar expression to Eq. (\ref{panalh}) with the following replacements $r_h \to r_c$, $\delta \to \hat{\delta}$, $\alpha \to \hat{\alpha}$ and $\beta \to \hat{\beta}$ where
\begin{align}
\hat{\delta}
=\sqrt{\hat{N} +1/12}\quad 
\hat{\alpha}=\sqrt{\frac{2}{\kappa r_c^2}}; \quad \hat{\beta}= \sqrt{\frac{2 R}{\kappa^2 r_c^2}}.
\end{align}
\subsection{$\langle \hat{\varphi}^2 \rangle_{ren}$}
\label{sec:Horizon ValueC}
Combining Secs. \ref{sec:HorizonValueA} and \ref{sec:HorizonValueB} we are in a position to write down an expression for $\langle \hat{\varphi}^2 \rangle_{ren}$ which is valid on both the event and cosmological horizons of a  lukewarm black hole, namely
 \begin{align}
\label{prenhc}
&\langle \hat{\varphi}^2 \rangle_{ren}=\frac{1}{8 \pi^2 r_a^2}\sum^{\infty}_{l=0}(2l+1)\beta'_l + \frac{1}{8\pi^2 r_0^2}\bigg[\frac{1}{12} +\frac{\beta'^2}{6\alpha'^4} \nonumber\\
&+\frac{d}{dx} \zeta\left(x,\textstyle{\frac{1}{2}} + i\delta'\right)\bigg|_{x=-1}+\frac{d}{dx} \zeta\left(x,\textstyle{\frac{1}{2}} - i\delta'\right)\bigg|_{x=-1}\nonumber\\
& -i \delta'\ln\left(\frac{\Gamma\left(x,\textstyle{\frac{1}{2}} + i\delta'\right)}{\Gamma\left(x,\textstyle{\frac{1}{2}} - i\delta'\right)}\right)+ N'\left(1  -\ln\left(\mu r_0 \right)\right) \bigg] \nonumber\\
&-\frac{ R(r_a) }{96 \pi^2 }+\frac{f'_a}{48\pi^2 r_a} +\frac{m^2}{16\pi^2},
\end{align}
with $N'= N$ or $\hat{N}$ when  $r_a=r_h$ or $r_c$  respectively and likewise for all other primed expressions.

\section{Plots of Results}
\label{sec:Plots}

\begin{figure}[htb!]\centering
\includegraphics[width=9cm]{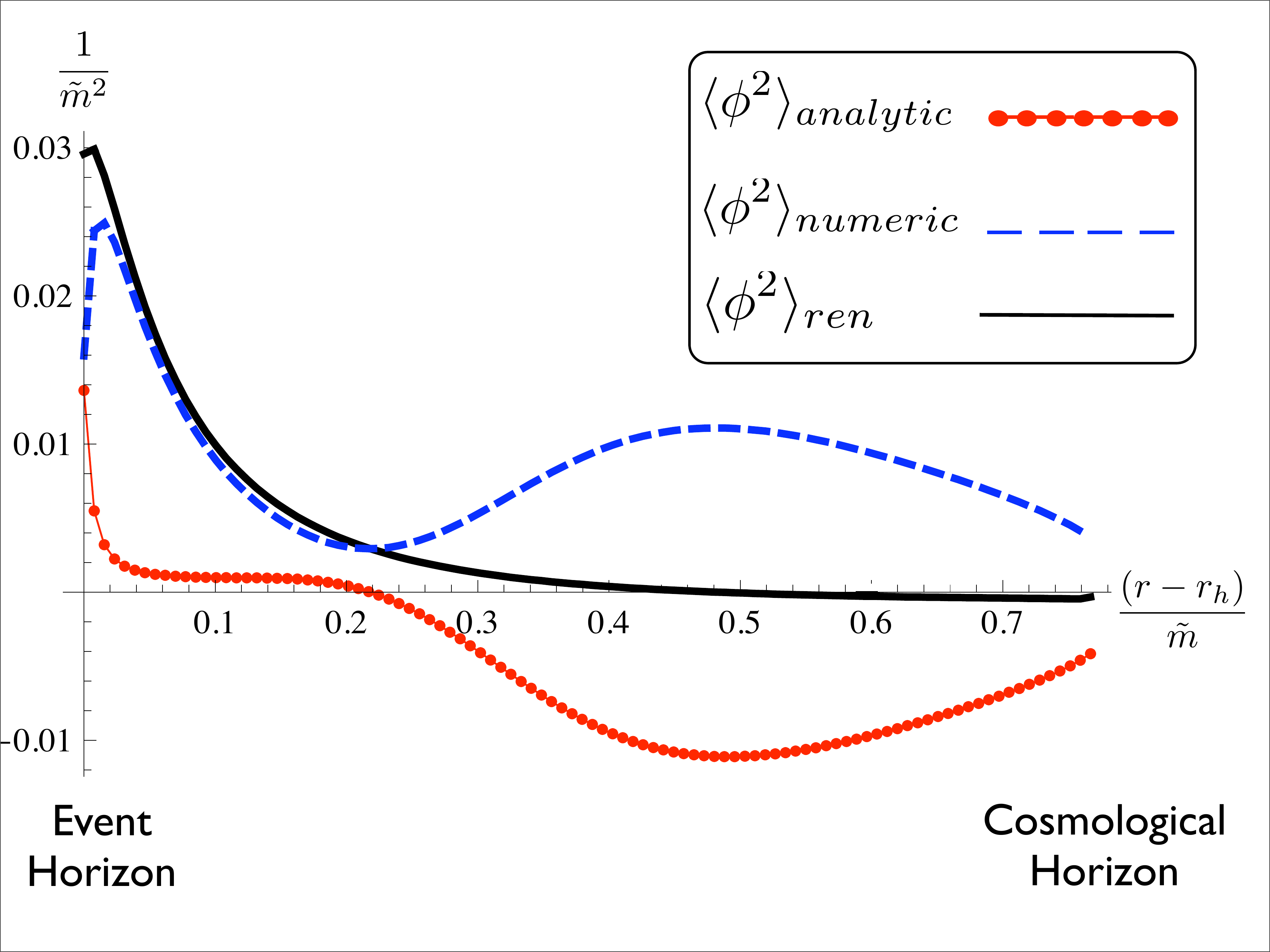}\caption{\label{fig:PhiRen}Plot of $\langle \hat{\varphi}^2 \rangle_{ren}$ and it components  for a massive conformally coupled scalar field with mass $m$ with the parameter set $M=Q=0.1\tilde{m}$ and $\Lambda= 3 \tilde{m}^{-2}$. It can be clearly seen that $\langle \hat{\varphi}^2 \rangle_{numeric}$, $\langle \hat{\varphi}^2 \rangle_{analytic}$ and hence  $\langle \hat{\varphi}^2 \rangle_{ren}$ are finite on both horizons }
\end{figure}

\begin{figure}[htb!]\centering
\includegraphics[width=9cm]{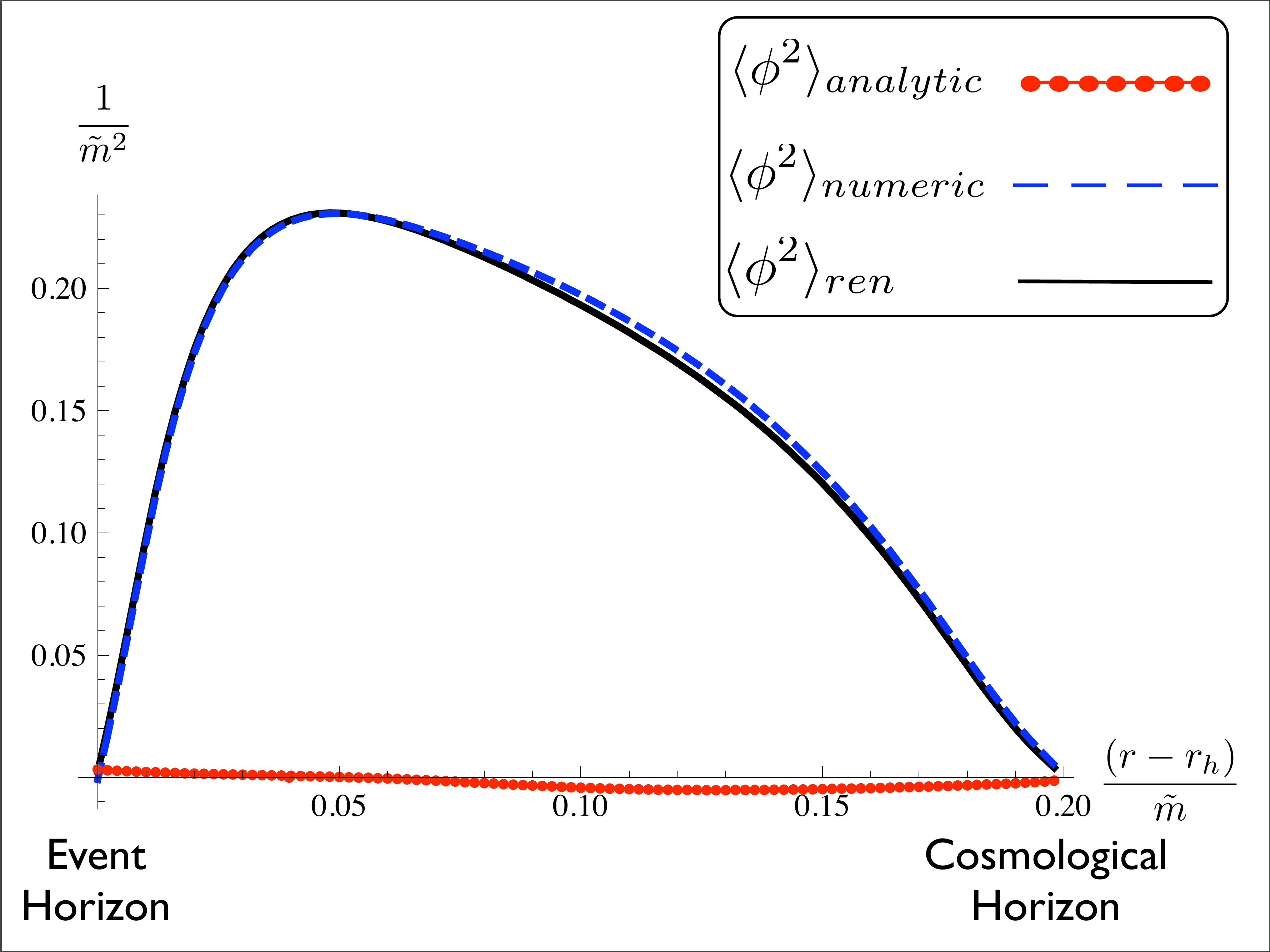}\caption{\label{fig:PhiRenEx}Plot of $\langle \hat{\varphi}^2 \rangle_{ren}$ and it components  for a massive, conformally coupled scalar field with mass $m$ with the parameter set $M=Q=0.24\tilde{m}$ and $\Lambda= 3 \tilde{m}^{-2}$. Here we are near the extremal limit $M=Q=0.25\tilde{m}$. In this case we see that $\langle \hat{\varphi}^2 \rangle_{numeric}$ dominates. Once more $\langle \hat{\varphi}^2 \rangle_{numeric}$, $\langle \hat{\varphi}^2 \rangle_{analytic}$ and hence  $\langle \hat{\varphi}^2 \rangle_{ren}$ are clearly finite on both horizons }
\end{figure}

We will now bring together the results from the previous sections. For the plots of $\langle \hat{\varphi}^2 \rangle_{ren}$ we calculate the quantities $\langle \hat{\varphi}^2 \rangle_{numeric}$ from Eq.~(\ref{numf}) and $\langle \hat{\varphi}^2 \rangle_{analytic}$ from Eq.~(\ref{analf}) at 100 grid points between the two horizons and take the value of $\langle \hat{\varphi}^2 \rangle_{ren}$ on the horizons to be given by Eq.~(\ref{prenhc}). For these calculations, unless otherwise stated, we express the dimensionful quantities ($Q$, $M$, $\Lambda$ and $r$) in units of the inverse of mass of the field $\tilde{m}$, which has dimension of length. In Figs. \ref{fig:PhiRen} and \ref{fig:PhiRenEx} we plot $\langle \hat{\varphi}^2 \rangle_{ren}$ for two different sets of parameters. 

\begin{figure}[htb!]
\includegraphics[width=9cm]{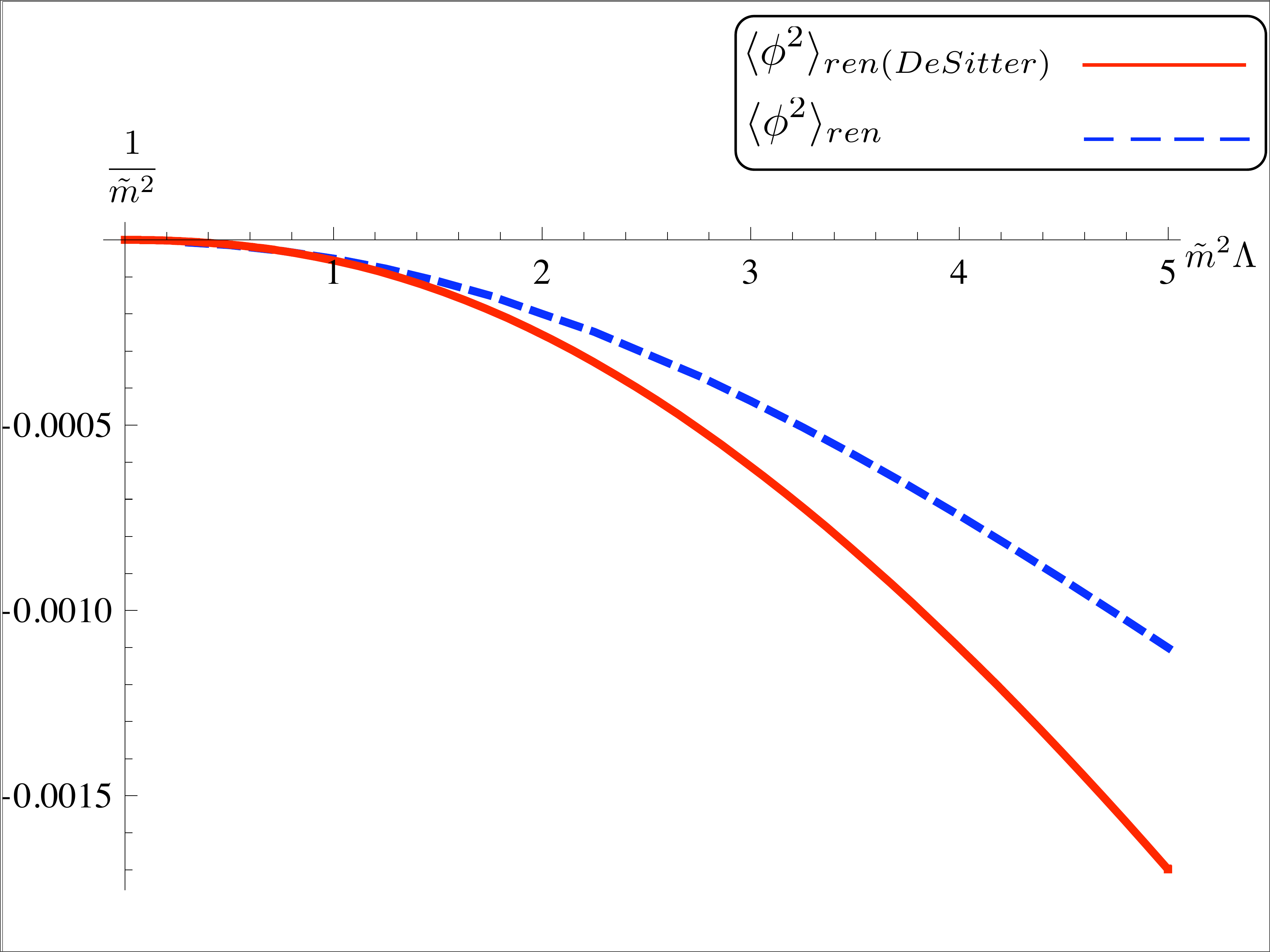}\caption{\label{fig:PhiCos}Plot of $\langle \hat{\varphi}^2 \rangle_{ren}$ for both the cosmological horizon of a lukewarm black hole and for a de Sitter space time as a function of the cosmological constant. As expected they appear to be in agreement as $\Lambda \to 0$.}
\end{figure}

In Fig.~\ref{fig:PhiCos} we plot the value of $\langle \hat{\varphi}^2 \rangle_{ren}$ on the cosmological horizon  of a lukewarm black hole for a massive conformally coupled scalar field with mass $m$ with the parameter set $M=Q=0.1\tilde{m}$ as a function of $\Lambda$. As $\Lambda \to 0$, the surface gravity (and hence the temperature) also tends to zero. In this limit, the cosmological horizon is pushed out to infinity. One would imagine then that the value of $\langle \hat{\varphi}^2 \rangle_{ren}$ on this horizon would be well approximated by the value of $\langle \hat{\varphi}^2 \rangle_{ren}$ in a deSitter space time at a temperature set by the black hole.
The line element for deSitter space time can be written as~\cite{BD}
\begin{equation}
ds^2 = \frac{\alpha^2}{\eta^2}\left[-d \eta^2 +\sum^{3}_{i=1} (d x^i)^2\right].
\end{equation}
where
\begin{align}
\eta=-\alpha e^{-{t}/{\alpha}};\quad \quad \quad
\alpha=\sqrt{\frac{3}{\Lambda}}.
\end{align}
So for temporal splitting the Green's function for this configuration is given by~\cite{BD}
\begin{align}
G(x,x')&=\frac{1}{16 \pi^2 \alpha^2}\left(\textstyle{\frac{1}{4}} -\nu^2\right)\sec\pi \nu\nonumber\\
&\qquad F\left(\textstyle{\frac{3}{2}}+\nu,\frac{3}{2} -\nu;2;1 +{\Delta\eta^2}/{(4 \eta \eta')}\right),
\end{align}
with
\begin{align}
\nu^2={\textstyle{\frac{9}{4}}}-12(m^2 R^{-1} +\xi)
\end{align}
and $F$ is a hypergeometric function.
It is now easily shown that for a quantum field in a de Sitter space-time at temperature $T={\kappa}/{(2\pi)}$
\begin{align}
&\langle \hat{\varphi}^2 \rangle_{ren}=\frac{m^2}{16 \pi^2} -\frac{1}{24 \pi^2 \alpha^2}  +\frac{1}{16 \pi^2} \left( m^2 +\left(\xi -\textstyle{\frac{1}{6}}\right)\frac{12}{\alpha^2}\right)\nonumber\\
&\times\left(\psi\left(\textstyle{\frac{3}{2}} +\nu\right) +\psi\left(\textstyle{\frac{3}{2}} -\nu\right)-1 -\ln(\mu^2 \alpha^2)\right).
\end{align}

Finally we examine the validity of the large field mass approximation to $\langle \hat{\varphi}^2 \rangle_{ren}$ obtained from the order $m^{-2}$ term in the DeWitt-Schwinger expansion \cite{Christensen}:
\begin{align}
\langle \hat{\varphi}^2 \rangle_{ren}& \approx\frac{1}{16 \pi^2 m^2}\left [\textstyle{\frac{1}{2}}\left(\xi -\textstyle{\frac{1}{6}}\right)^2 R^2 -\frac{1}{6}\left(\xi -\textstyle{\frac{1}{5}}\right)^2  R\right.\nonumber\\
&\left.+ \textstyle{\frac{1}{280}}(R_{\alpha \beta \gamma \delta} R^{\alpha \beta \gamma \delta} -R_{\alpha \beta} R^{\alpha \beta} ) \right],
\end{align}
where $R_{\alpha \beta \gamma \delta} $ is the Riemann tensor and $R_{\alpha \beta}$ is the Ricci tensor.
In a Reissner-N\"{o}rdstrom-de Sitter spacetime this becomes
\begin{align}
\label{DSapprox}
&\langle \hat{\varphi}^2 \rangle_{ren} \approx\frac{1}{16 \pi^2 m^2}\left [\textstyle{\frac{1}{2}}\left(\xi -\textstyle{\frac{1}{6}}\right)^2 R^2 -\frac{1}{6}\left(\xi -\textstyle{\frac{1}{5}}\right)^2  R\right.\nonumber\\
&\ \left.+ \textstyle{\frac{1}{280}}\left(({ 52 Q^4-96M Q r +48 M^2 r^2})/{r^8} -\frac{4}{3}\Lambda\right)\right].
\end{align}
To compare this with our numerical results, we give all our dimensionful quantities in units of $L \equiv\sqrt{3/\Lambda}$. This allows us to vary the mass of the field in our results.
Our numerical calculations then show that Eq.~(\ref{DSapprox}) is a reasonable approximation for, but only for, very large values of the field mass $m$ ($m \approx 20L^{-1}$ when $M=Q=0.1 L$). This is consistent with the calculations of Anderson~\cite{Anderson:1990jh} for a Reissner-N\"{o}rdstrom space time.

\section{Conclusions}
\label{sec:Conclusions}
In this paper we have considered a scalar quantum field in a Hartle-Hawking state in a space-time with two horizons at equal temperatures. We introduced a new uniform approximation to $C_{n l}p_{n l}q_{n l}$ and have used this approximation to calculate $\langle \hat{\varphi}^2 \rangle_{ren}$ on a lukewarm Reissner-N\"{o}rdstrom-de Sitter space time. The uniformity of this approximation allows us to calculate $\langle \hat{\varphi}^2 \rangle_{ren}$ on both the event and cosmological horizon dealing always with finite quantities. 
We have also shown that for small values of the cosmological constant, $\langle \hat{\varphi}^2 \rangle_{ren}$ on the cosmological horizon approaches $\langle \hat{\varphi}^2 \rangle_{ren}$ for a de Sitter spacetime at temperature $T=\kappa/2\pi$.

 While other uniform approximations \cite{Tom} have been used to prove the regularity of $\langle \hat{\varphi}^2 \rangle_{ren}$ on the the black hole horizons \cite{Winstanley:2007}, they fail to give explicit horizon values for $\langle \hat{\varphi}^2 \rangle_{ren}$. Also they cannot be used to investigate the regularity of $\langle T_{\mu \nu} \rangle_{ren}$  on the black-hole horizons.
Our new approximation overcomes these shortcomings which, as we have just demonstrated, allows us to obtain explicitly finite and easily calculable values of $\langle \hat{\varphi}^2 \rangle_{ren}$  on the two horizons and provides a route for us  to extend our consideration to $\langle T_{\mu \nu} \rangle_{ren}$~\cite{BOWY}.
 
Finally we note that throughout this project we been able to perform all our numerical calculations using \textsl{Mathematica}~\cite{Mathematica}. This greatly expedites the code development process and enables
easy modification to other spacetimes.

\acknowledgments
We would like to thank Daniel Foran for his invaluable contribution to the early stages of this project. We also wish to thank Elizabeth Winstanley for many helpful conversations. The work of CB is supported by the Irish Research Council for Science, Engineering and Technology, funded by the National Development Plan.

\appendix
\section{Near Horizon Expansions}
\label{Ap:NearHorizon}
In this Appendix we describe the near horizon expansions needed for various calculations throughout this paper. We begin with
\begin{align}
\label{fexpan}
f\approx f'_0\epsilon +\frac{f''_0}{2} \epsilon^2 +O(\epsilon^3),
\end{align}
with $r_0$ denoting the position of the horizon, $\epsilon = r-r_0$ and $f'_0\equiv f'(r_0)$, etc.
From Eq.~(\ref{fexpan})  we can deduce
\begin{align*}
\zeta(r) &=\int_{r_0}^r \frac{\mathrm{d}r'}{r' f(r')^{1/2}}=\frac{2}{ r_0^2 {f'_0}^{1/2}} \epsilon^{1/2} +O(\epsilon^{3/2}),\\
A(r) &= \frac{\zeta(r)}{rf(r)^{1/2}}= \frac{2}{r_0^2 f'_0 } +O(\epsilon).
\end{align*}
We also need to calculate expansions for the radial geodesic distance $s$ from $r_0$ to $r$. First we have
\begin{equation*}
s_r = f(r)^{-1/2} = \frac{1}{{f'_0}^{1/2}} {\epsilon}^{-1/2} - \frac{f''_0}{4 {f'_0}^{3/2}} \epsilon^{1/2} + O(\epsilon^{3/2}) ,
\end{equation*}
and integrating yields
\begin{equation*}
s = \frac{2}{{f'_0}^{1/2}} {\epsilon}^{1/2}- \frac{f''_0}{6 f_ 0^{'3/2}} \epsilon^{3/2} + O(\epsilon^{5/2}).
\end{equation*}
The combinations appearing in the Christensen subtraction terms are then
\begin{align*}
\frac{1}{4 \pi^2 s^2} &=\frac{f'_0}{16 \pi^2 \epsilon} + \frac{f''_0}{96 \pi^2} +O(\epsilon),\\
R^{\alpha\beta}s_{\alpha}s_{\beta}&=R^{rr}s_{r}{}^{\!2}= \left[-\frac{\left(f''_0 r_0 + 2f'_0 \right) }{2r_0}  f'_0\epsilon +O(\epsilon^2)\right]s_{r}{}^{\!2}\\
&=-\frac{\left(f''_0 r_0 + 2f'_0 \right) }{2r_0}  +O(\epsilon).
\end{align*}

\section{Recursion Relations}
\label{Ap:RecursionRelations}
 In this Appendix we derive the recursion relations needed for the initial conditions described in Sec.~\ref{sec:NumericalCalculations}. Firstly we rewrite Eq.~(\ref{mode}) in the form
\begin{align}
\label{mode2}
 &r^2 f\frac{\mathrm{d}}{\mathrm{d}r}\left(r^2 f \frac{\mathrm{d}S }{\mathrm{d}r} \right)\nonumber\\
&-\left(n^2 \kappa^2  +r^2 f({l(l+1)} +(m^2 +\xi R)r^2) \right) S=0.
\end{align}
We define the following expansions
\begin{align*}
&r^2 f= \frac{\Lambda}{3} (r_c-r)(r-r_h)(r-r_n)(r-r_i),\nonumber\\
&(r^2 f)^2=\sum_{j=2}^{8} A_j (r_c -r)^j,\nonumber\\
&(r^2 f)\frac{\mathrm{d}(r^2 f)}{\mathrm{d}r}=\sum_{j=1}^{7} B_j (r_c -r)^j,\nonumber\\
&n^2 \kappa^2  +r^2 f({l(l+1)} +(m^2 +\xi R)r^2)=\sum_{j=0}^{6} V_j (r_c -r)^j 
\end{align*}
where $A_j$, $B_j$ and $V_j$ are easily determined.
Then inserting the series~(\ref{series q}) into Eq.~(\ref{mode2}) and we obtain the following seven-term recursion relation for the $b_i$ coefficients:
\begin{align*}
&b_i =-\Bigl( \sum_{j=3}^{8} (i +2 +n/2 -j)(i +1 +n/2 -j)A_j b_{i-j+2}\nonumber\\
&  + \sum_{j=2}^{7} (i +1 +n/2 -j)B_j b_{i-j +1} 
- \sum_{j=1}^{6}V_j b_{i-j}\Bigr) \big/{G_i(l,n)} ,
\end{align*}
with $G_i(l,n)= (i +n/2)((i+n/2 -1)A_2 +B_1 -V_0)$ and we normalise by taking $b_0 =1/\sqrt{\kappa r_c^2}$.

We now turn to calculating the recursion relation needed to determine the coefficients in our  Jaffe-like solution  for $p_{n l}$.
We first change variable in Eq.~(\ref{mode}) to $x=r-r_i$. We then perform the following transformation:
\begin{equation*}
x \to u=\frac{x-x_h}{x} \quad \textrm{and} \quad\ U(x) \to x_h^{n/2}(1-u)^6 Y(u),
\end{equation*}
where $x_h=r_h-r_i$.
 Eq.~(\ref{mode}) then takes the form:
\begin{align}
\label{equ}
u F(u) \frac{\mathrm{d}}{\mathrm{d}u}&\left( u F(u) \frac{\mathrm{d}}{\mathrm{d}u}(1-u)^6 Y(u)\right)  \nonumber \\
&\qquad -V_{ln}(u)(1-u)^4 Y(u)=0.
\end{align}
where
\begin{equation*}
\label{F}
F(u) =(x- x_n)(x_c -x) = \Bigl(\frac{x_h}{1-u}-x_n\Bigr)\Bigl(x_c-{\frac{x_h}{1-u}}\Bigr)
\end{equation*}
and
\begin{align*}
V_{ln}(u)& =\frac{n^2 \kappa^2(1-u)^2}{x_h^2} \Bigl(\frac{x_h}{1-u} +r_i\Bigr)^4 \nonumber\\
&+u F(u) \left(l(l+1) +(m^2 +\xi R)\Bigl({\frac{x_h}{1-u}} +r_i\Bigr)^2\right)
\end{align*}
with $x_n$ and $x_c$ denoting the position of the corresponding horizons in the variable $x$.

We now try a series solution of the form
\begin{equation*}
Y(u)=\sum^{\infty}_{i=0} c_i u^{i+n/2}.
\end{equation*}
We define the following expansions:
\begin{align}
\label{expans}
&(u F)^2(1-u)^6= \sum^{8}_{j=2} \hat{A}_j u^j\nonumber\\
&(u F)\frac{d}{du}(u F)(1-u)^6 
-12(1-u)^5(uF)^2= \sum^{7}_{j=1} \hat{B}_j u^j,\nonumber\\
&30(uF)^2(1-u)^4
-  6(u F)\frac{d}{du}(u F)(1-u)^5=\sum^{6}_{j=1} \hat{C}_j u^j,\nonumber\\
&V_{l,n}(u)(1-u)^4 = \sum^{6}_{j=0} \hat{V}_j u^j\nonumber
\end{align}
where the coefficients $\hat{A}_j$, $\hat{B}_j$, $\hat{C}_j$, and $\hat{V}_j$ are easily determined. 
Inserting the above expansions into Eq.~(\ref{equ}), we obtain the following seven-term recurrence relation for the series coefficients $c_i$:
\begin{align*}
&c_i =-\Bigl(\sum_{j=3}^{8} (i-j +n/2 +2)(i-j +n/2 +1)\hat{A}_j c_{i-j +2}\\
&+\sum_{j=2}^{7} (i-j +n/2 +1) \hat{B}_j c_{i-j +1}\\
&+\sum_{j=1}^{6} \hat{C}_j c_{i-j}- \sum_{j=1}^{6} \hat{V}_j c_{i-j}\Bigr)\big/\hat{G}_i (n,l)
\end{align*}
with
$
\hat{G}_i(l,n)= (i+n/2)( (i+n/2 -1)\hat{A}_2 +\hat{B}_1) - \hat{V}_0,
$
and we normalise by taking $c_0=1/\sqrt{\kappa r_h^2}$.

\section{Calculation of $\beta_l$}
 \label{Ap:Beta}
To calculate the constants   $\beta_l$ we make use of the Wronskian of Eq.~(\ref{mode})
\begin{equation*}
p_l(r)\frac{\mathrm{d}}{\mathrm{d}r}q_l(r)-q_l(r)\frac{\mathrm{d}}{\mathrm{d}r}p_l(r)=-\frac{1}{r^2f}.
\end{equation*}
We will calculate $\beta_l$ for the cosmological horizon, $\hat{\beta}_l$; the calculation for the event horizon follows
identical lines.
Dividing the Wronskian by $q_l(r)^2$ and integrating yields an integral representation for
$p_l(\eta)$
\begin{equation}
\label{b1}
p_l(r)=q_l(r)\int^{r}_{r
_h} \frac{\mathrm{d}r'}{r'^2f(r') q_l(r')^2}
\end{equation}
with $r_h$ being the location of the inner boundary.

Near the cosmological horizon we have a series solution for $q_l(r)$
\begin{equation}
\label{p1}
q_l(r)=a +b_l(r_c-r)+O((r_c- r)^2)
\end{equation}
where $a=1/\sqrt{\kappa r_c^2}$ and the $b_l$ are constants.
Hence we can rewrite Eq.~(\ref{b1}) in the form
\begin{align}
\label{eq:b3}
p_l(r)&=q_l(r)\int^{r}_{r_h} \frac{\mathrm{d}r'}{r'^2f(r')}\bigg(\frac{1}{q_l(r')^2}-\frac{1}{a^2}\bigg)\nonumber\\
&\qquad\qquad+ \frac{q_l(r)}{a^2}\int^{r}_{r_h} \frac{\mathrm{d}r'}{r'^2f(r')}.
\end{align}
Now the first integral converges as $r\to r_c$ while the second integral can be computed exactly as
\begin{align*}
&\int \frac{\mathrm{d}r}{r^2f}= \frac{1}{\kappa_h r_h^2} \ln(r-r_h)+\frac{1}{2\kappa_ir_i^2} \ln(r-r_i)\nonumber\\
&\qquad +\frac{1}{2\kappa_n r_n^2} \ln(r-r_n)-\frac{1}{2\kappa_c r_c^2} \ln(r_c-r)\equiv I(r).
\end{align*}

Inspection of this expression reveals that we pick up a divergence when either of the endpoints of integration coincides with a horizon. This gives us the divergence we want on the cosmological horizon. However, we also  pick up an unwanted divergence
at $r_a=r_h$; this divergence is not real as it is cancelled by a similar divergence in the first integral. To show this cancellation explicitly we integrate Eq.~(\ref{eq:b3}) by parts to obtain:
\begin{align*}
&\int^{r_c}_{r_h} \frac{\mathrm{d}r'}{r'^2f(r')}\bigg(\frac{1}{q_l(r')^2}-\frac{1}{a^2}\bigg) + \frac{1}{a^2}\int^{r}_{r_h} \frac{\mathrm{d}r'}{r'^2f(r')} \\
&=\bigg[\frac{I(r')}{q_l(r')^2}-\frac{I(r')}{a^2}\bigg]^r_{r_h}+\int^{r}_{r_h} \frac{2 q'_l(r')}{q_l^3(r')}I(r') \mathrm{d}r'+\bigg[\frac{I(r')}{a^2}\bigg]^r_{r_h}\\
&=\frac{I(r)}{q_l(r)^2} +\int^{r}_{rh} \frac{2 q'_l(r')}{q_l^3(r')}I(r') \mathrm{d}r',
\end{align*}
where we have used  $I(r)/q^2(r)\to 0$ as $r\to r_h$.

We may now obtain an asymptotic expansion for $p_l(r)$ as $r \to r_c$ using (\ref{eq:b3}),
\begin{align}
\label{asym1}
&p_l(r) =\frac{1}{\sqrt{\kappa r_c^2}} \int^{r_c}_{r_h} \frac{2 q'_l(r')}{q_l^3(r')}I(r') \mathrm{d}r'-\frac{1}{2 \sqrt{\kappa r_c^2}} \ln(r_c-r)\nonumber\\
&+\sqrt{\kappa r_c^2}\bigg[\frac{1}{\kappa_h r_h^2} \ln(r_c-r_h)+\frac{1}{2\kappa_ir_i^2} \ln(r_c-r_i)\nonumber\\
&+ \frac{1}{2\kappa_n r_n^2} \ln(r_c-r_n) \bigg]
+ O((r_c-r) \ln(r_c-r) ).
\end{align}

Alternatively, we may expand the cosmological horizon equivalent of relation (\ref{qexpan}) 
\begin{equation*}
p_{0l}(r)=\hat{P}_{0}(r) +\hat{\beta}_l q_{0l}(r) + \hat{R}_l(r),
\end{equation*}
about $r=r_c$ to obtain \cite{gradriz}
\begin{align}
\label{asym2}
p_l(r)&= \frac{1}{\sqrt{\kappa r_c^2}} \Bigl(-\ln(\hat{k})+\frac{1}{2}\ln(2\kappa r_c^2) -\frac{1}{2}\ln(r_c-r)  \nonumber\\
&-\gamma+\hat{\beta}_l\Big) +O((r_c-r) \ln(r_c-r) ).
\end{align}
Comparison of Eqs. (\ref{asym1}) and (\ref{asym2}) provides us with an expression for the $\hat{\beta}_l$ which is amenable to numerical calculation:
\begin{align*}
&\hat{\beta}_l =\int^{r_c}_{r_h} \frac{2 q'_l(r')}{q_l^3(r')}I(r') \mathrm{d}r' +\kappa r_c^2\bigg[\frac{1}{\kappa_h r_h^2} \ln(r_c-r_h)\nonumber\\
&\quad+\frac{1}{2\kappa_ir_i^2} \ln(r_c-r_i)+\frac{1}{2\kappa_n r_n^2} \ln(r_c-r_n) \bigg]\nonumber\\
&\qquad+\ln(\hat{k})+\gamma-\frac{1}{2}\ln(2\kappa r_c^2) .
\end{align*}

\section{Calculations needed for $\langle\hat{\varphi}^2\rangle_{analytic}$}
 \label{Ap:Analytic}
Here we present the details of the calculation of
$
\sum^{\infty}_{l=0}F_0(l)
$ and $\int_{0}^{\infty}  F_1(x) \>\mathrm{d}x$ with $F_0$ and $F_1$ defined by Eqs.~(\ref{defF0}) 
and (\ref{defF1}).

 Firstly we consider the sum, that is
\begin{equation}
\label{F0}
\sum^{\infty}_{l=0}\left\{(2l+1)[-\ln(k)+\psi(l+1)]+\frac{N}{l+1}\right\}.
\end{equation}
It is convenient to introduce a large $l$ cut off $L$,  evaluate each individual sum separately and show that the terms in the individual sums which are divergent as $L\to\infty$ cancel, leaving a finite analytic remainder.
We start by considering the sum
\begin{equation}
\label{s1}
\sum^{L}_{l=0} (2l+1)\ln(k)
\end{equation}
To do this we begin by writing $k^2$ as
\begin{equation*}
k^2 =\left(\left(l+\textstyle{\frac{1}{2}}\right) + i \delta\right)\left(\left(l+\textstyle{\frac{1}{2}} \right)- i \delta\right)
\end{equation*}
where $\delta = \sqrt{1/12 +N}$,
then  we can rewrite the sum (\ref{s1}) as
\begin{equation*}
\sum^{L}_{l=0} (l+1/2)\Bigl[\ln \left(\left(l+\textstyle{\frac{1}{2}}\right) + i \delta\right)+\ln \left(\left(l+\textstyle{\frac{1}{2}}\right) - i \delta\right)\Bigr].
\end{equation*}

We now consider 
$
\zeta(z,L+1+q) -\zeta(z,q)
$
where $\zeta$ is the generalised Riemann Zeta Function.
We can reexpress this in the form \cite{gradriz}
\begin{align*}
&\sum^{\infty}_{k=0} \frac{1}{(L+1+q +k)^z} -\sum^{\infty}_{k=0} \frac{1}{(q +k)^z}\\
&=\sum^{\infty}_{k'=L+1} \frac{1}{(q +k')^z}-\sum^{\infty}_{k=0} \frac{1}{(q +k)^z}=-\sum^{L}_{k=0} \frac{1}{(q +k)^z}\\
&=-\sum^{L}_{k=0} e^{-z \ln(q+k)} .
\end{align*}
Differentiating with respect to $z$ gives:
\begin{equation*}
\frac{\mathrm{d}}{\mathrm{d}z}\bigl(\zeta(z,L+1+q) -\zeta(z,q)\bigr)=\sum^{L}_{k=0} (q+k)^{-z} \ln(q+k)
\end{equation*}
and so setting $z=-1$ we have
\begin{equation}
\label{d-1}
\frac{\mathrm{d}}{\mathrm{d}z}\bigl(\zeta(z,L+1+q) -\zeta(z,q)\bigr)\bigg|_{z=-1} =\sum^{L}_{k=0} (q+k)\ln(q+k),
\end{equation}
while setting $z=0$ gives
\begin{equation}
\label{d0}
\frac{\mathrm{d}}{\mathrm{d}z}\bigl(\zeta(z,L+1+q) -\zeta(z,q)\bigr)\bigg|_{z=0}=\sum^{L}_{k=0} \ln(q+k).
\end{equation}
Combining Eqs.~(\ref{d-1}) and ({\ref{d0}),
and
using the identity~\cite{gradriz}
\begin{equation*}
\frac{\mathrm{d}}{\mathrm{d}x} \zeta(x,\delta)\bigg|_{x=0}=\ln\bigl(\Gamma(\delta)\bigr) -\frac{1}{2}\ln(2\pi)
\end{equation*}
yields the result
\begin{align}
\label{ksum}
&\sum^{L}_{l=0} (2l+1)\ln(k)=\frac{\mathrm{d}}{\mathrm{d}x} \bigl( \zeta\left(x,i L+\textstyle{\frac{3}{2}}+ \delta \right)-\zeta\left(x,\textstyle{\frac{1}{2}} + i\delta\right)\nonumber \\
&\qquad+ \zeta\left(x,L+\textstyle{\frac{3}{2}}-i \delta\right)- \zeta\left(x,\textstyle{\frac{1}{2}} - i\delta\right)\Bigr)\bigg|_{x=-1}\nonumber\\
&+i \delta\bigg[\ln\left(\frac{\Gamma\left(\textstyle{\frac{1}{2}} + i\delta\right)}{\Gamma\left(\textstyle{\frac{1}{2}} - i\delta\right)}\right)-\ln\left(\frac{\Gamma\left(L+\textstyle{\frac{3}{2}}+i \delta\right)}{\Gamma\left(L+\textstyle{\frac{3}{2}}-i \delta\right)}\right)\bigg].
\end{align}
To evaluate the second term in Eq.~(\ref{F0}) we consider
\begin{equation*}
(z^2-1)\sum^{L}_{l=0} \psi(l+1) z^{2l} .
\end{equation*}
Multiplying out gives
\begin{align}
\label{psi}
&\sum^{L}_{l=0} \psi(l+1) z^{2(l+1)}-\sum^{L}_{l=0} \psi(l+1) z^{2l}\nonumber \\ 
&=\sum^{L}_{l=1} (\psi(l) - \psi(l+1)) z^{2l}  -\psi(1) + \psi(L+1)z^{2(L+1)}.\nonumber \\
\end{align}
Now we have the identities~\cite{gradriz}
\begin{equation*}
\psi(l) - \psi(l+1)=\frac{1}{l} \quad \textrm{and}\quad \psi(1)=-\gamma.
\end{equation*}
Inserting these into Eq.~(\ref{psi}) and rearranging gives
\begin{equation*}
\sum^{L}_{l=0} \psi(l+1) z^{2l+1} =\frac{z}{z^2-1}\bigg[\gamma + \psi(L+2)z^{2(L+1)}-\sum^{L+1}_{l=1} \frac{ z^{2l}}{l}  \bigg].
\end{equation*}
Now we expand the right-hand side about $z=1$ and simplify to give
\begin{align*}
& (L+1)\frac{z}{z+1} \bigg[  2 \left(\psi(L+2)-1\right)  \\
&-\bigl((L+1)-(2L+1)\psi(L+2)\bigr)(z-1) + O\left((z-1)^2\right)\bigg],
\end{align*}
so differentiating this expression with respect to $z$ and then setting $z=1$ we find
\begin{align}
\label{psisum}
\sum^{L}_{l=0} &(2l+1)\psi(l+1)\nonumber\\
&= -\frac{1}{2} (L+1)(L+2) +(L+1)^2\psi(L+2).
\end{align}
The final sum may be performed immediately~\cite{gradriz}
\begin{equation}
\label{Nsum}
\sum^{n}_{l=0} \frac{N}{l+1} = N(\gamma + \psi(n+2)).
\end{equation}
Combining Eqs.~(\ref{ksum}), (\ref{psisum}), (\ref{Nsum}) gives the result
\begin{widetext}
\begin{align}
\label{eq:lim}
&\sum^{L}_{l=0}F_0(l)=-\frac{\mathrm{d}}{\mathrm{d}x}\bigl(  \zeta\left(x,L+\textstyle{\frac{3}{2}}+i \delta\right)+
\zeta\left(x,L+\textstyle{\frac{3}{2}}-i \delta\right)\bigr) \bigg|_{x=-1} -\textstyle{\frac{1}{2}} (L+1)(L+2)+\bigl((L+1)^2+ N) \psi(L+2) \nonumber\\
& 
+ i \delta\ln\left(\frac{\Gamma\left(L+\textstyle{\frac{3}{2}}+i \delta\right)}{\Gamma\left(L+\textstyle{\frac{3}{2}}-i \delta\right)}\right) + \frac{\mathrm{d}}{\mathrm{d}x}\left(
\zeta\left(x,\textstyle{\frac{1}{2}} + i\delta\right) + \zeta\left(x,\textstyle{\frac{1}{2}} - i\delta\right)\right)\bigg|_{x=-1} - i \delta\ln\left(\frac{\Gamma\left(\textstyle{\frac{1}{2}} + i\delta\right)}{\Gamma\left(\textstyle{\frac{1}{2}} - i\delta\right)}\right)+N \gamma
\end{align}
\end{widetext}
We now consider the limit as $L\to \infty$; for large $z$~\cite{dlmf}
\begin{align}
\label{zexpan}
\frac{\mathrm{d}}{\mathrm{d}x} \zeta(x,z)\bigg|_{x=-1}&=\textstyle{\frac{1}{2} z^2 \ln z -\frac{1}{4}z^2 -\frac{1}{2}z\ln z +\frac{1}{12} \ln z} \nonumber\\
& \textstyle{+\frac{1}{12}}+O\left({z}^{-1}\right),
\end{align}
\begin{equation}
\label{gexpan}
\ln\Gamma(z)= \textstyle{\left(z-\frac{1}{2}\right)\ln z -z -\frac{1}{2}\ln(2\pi) }+O\left({z}^{-1}\right),
\end{equation}
\begin{equation}
\label{psiexpan}
\psi(z) =\textstyle{\ln z  -\frac{1}{2} z^{-1}-\frac{1}{12 } z^{-2}+O\left(z^{-3}\right)}.
\end{equation}
Using Eqs.~(\ref{zexpan}), (\ref{gexpan}), (\ref{psiexpan}) and the definition of $\delta$ it is straightforward to show that in the limit as $L\to\infty$ the 
$L$-dependent terms in Eq.~(\ref{eq:lim}) tend to $N+1/12$.

Now examining the integral term, we have \cite{Watson}, \cite{gradriz}
\begin{align*}
\int_{0}^{\infty}  &F_1(x) \> \mathrm{d} x =\\
&\bigg[- \frac{\beta^2}{6 \alpha^3}K_1(\alpha x)+\frac{N}{\alpha^2}{K_0(\alpha x)} + N \mathrm{Ei}\left(-\frac{\alpha x}{\sqrt{2}}\right)\bigg]_{0}^{\infty}\\
\end{align*}
where $\mathrm{Ei}$ is the exponential integral function.  Each term decays exponentially in $x$ as $x\to \infty$~\cite{gradriz}. At the lower limit each term diverges but by expanding about $x=0$~\cite{gradriz}, we see that the combination yields a finite limit as $x\to 0$ giving 
 \begin{equation}
\int_{0}^{\infty}  F_1(x) = \frac{\beta^2}{6\alpha^4} -N\log(\sqrt{2}).
\end{equation}

\end{document}